\documentclass[aps, prd, twocolumn, lengthcheck, superscriptaddress, showpacs, letterpaper, nofootinbib]{revtex4-1}

\usepackage{epsfig}
\usepackage{amsmath,amssymb}
\usepackage{graphicx,bm}
\usepackage{subfigure}
\usepackage{slashed}
\usepackage{dcolumn}
\usepackage{epstopdf}
\usepackage{ulem} %% for strike-through
\usepackage[usenames]{color}
\usepackage{mathrsfs}

\allowdisplaybreaks

\begin{document}

\title{Chiral soliton lattice effect on baryonic matter from skyrmion crystal model}

\author{Mamiya Kawaguchi}

\email{mkawaguchi@hken.phys.nagoya-u.ac.jp}
\address{Department of Physics, Nagoya University, Nagoya 464-8602, Japan.}

\author{Yong-Liang Ma}
\email{yongliangma@jlu.edu.cn}
\affiliation{Center for Theoretical Physics and College of Physics, Jilin University, Changchun,
130012, China}

\author{Shinya Matsuzaki}
\email{synya@hken.phys.nagoya-u.ac.jp}
\address{Department of Physics, Nagoya University, Nagoya 464-8602, Japan.}
\affiliation{Center for Theoretical Physics and College of Physics, Jilin University, Changchun,
130012, China}

\date{\today}
%%%%%%%%%%%%%%%%%%%%%%%%%%%%%%%
\begin{abstract}
The chiral soliton lattice (CSL) has been studied in condensed-matter system such as chiral magnets, which arises as a parity-violating topological soliton.
In hadron physics, various attempts have also been made to apply the idea of CSL to baryonic matter.
In this work, we explore the CSL effects on the baryonic matter based on the skrymion crystal approach.
It is found that the CSL causes an inverse catalysis for the topology change in the dense baryonic matter.
Furthermore, we observe that the CSL makes the single-baryon shape deformed to be highly oscillating as the frequency of the CSL gets larger, which leads to the enhancement of a baryon energy. Of interest is that in a high density region, the CSL
goes away due to the topology change in the baryonic matter.
What we find here might deepen our understanding of dense matter systems as well as compact stars.
\end{abstract}


\maketitle

%%%%%%%%%%%%%%%%%%%%%%%%%%%%%%%%%%%%%%%%%%%%%%%%%%%%%%%%%%%%%%%%%%%%%%%

\section{introduction}

\label{sec:intro}

The phase structure of QCD is a major subject for the strong interaction physics. At low temperature and/or high baryon number density region, many parts of the phase diagram remain
not yet clarified due to the nonperturbative nature of QCD.  In particular, the high density region of nuclear matter cannot be accessed so far from lattice simulation based on the fundamental QCD due to the notorious sign problem. One would therefore resort the analysis of the nuclear matter in high density region to effective models of QCD~\cite{Holt:2014hma,Krewald:2012zza,MRBook}.
Since these effective models often posses topological objects, say a soliton which will be focused on throughout the present paper, by considering some analogy between dense nuclear matter and condensed matter physics,
it would be reasonable to expect that such approaches give a new insight for understanding the phase structure of QCD.

In the condensed matter systems, a periodic and parity-violating topological soliton called chiral soliton lattice (CSL) has been studied in chiral magnets~\cite{Dzyaloshinsky:1964}
and its structure  was experimentally observed in the recent investigations for chiral magnets~\cite{Togawa:2012}.
Given the fact that the CSL structure was discovered in condense matter physics,
some attempts have been made to adapt the idea of CSL to QCD in the context of the high energy physics~\cite{Brauner:2016pko}. In particular, a single domain wall solution of the CSL, which is called  pion domain wall regarded as pion condensation, has been addressed in searching for new aspects of a dense nuclear matter~\cite{Hatsuda:1986nq,Son:2007ny,Eto:2012qd}.
However, the CSL effect on properties of baryonic matter such as
the deformation of the baryonic matter structure has not been fully examined in the hadron physics, so still lots of rooms for this avenue are left.

In this paper, we analyze the properties of nuclear matter including the CSL
structure based on the skrymion crystal model which provides a novel approach to dense baryonic matter in which the nuclear matter and the medium-modified properties of hadrons can be accessed in a unified way~\cite{ParkVento,Ma:2016gdd}. In this approach, the dense baryonic matter is accessed by putting the skyrmions onto the crystal lattice and regarding the skyrmion matter as baryonic matter.
What is observed in this crystal (but not in other approaches) is the topology change,
i.e., the crystal lattice made of skyrmions is deformed into another crystal lattice made of half-skyrmions. And, it is found that such a topology change has significant effect of the equation of state of dense nuclear matter and consequently the properties of compact stars~\cite{PKLMR,MLPR,Ma:2018xjw}.

We find that the presence of the CSL plays the role of the enhancement
of the skyrmion crystal energy and an inverse catalyzer for the topology change on the baryonic matter.
In addition,  the CSL distorts the skyrmion crystal configuration.
In a low density region (before topology change), the single-skymion in the crystal is deformed to be intense objects with a definite frequency due to the periodicity of the CSL. We also observe that
as the periodicity of the CSL gets shorter, the single-skymon is more oscillating,
which leads to the enhancement of a single-baryon energy.
On the other hand, in a high density region (after topology change), the CSL actually goes away in relation to the topology change on the baryonic matter. Those findings might be relevant also to the understanding of  condensed-matter systems as well as compact stars.

This paper is organized as follows:
In sec.~\ref{sec:model} we introduce the basic setup in studying  the CSL effects on the skyrmion crystal.
In sec.~\ref{num_res} we show our numerical analysis for the dependence of the CSL on the skyrmion crystal and some related phenomena such as the topology change on the baryonic matter, the deformation of skyrmoin crystal structure and single baryon shape. A summary is given in sec.~\ref{sum} .

\section{CSL in the Skyrmion crystal }

\label{sec:model}

\subsection{Skyrmion crystal}

In the present work, we employ
the following Skyrme model Lagrangian with the pion mass term included
\cite{Skyrme:1962vh},
\begin{eqnarray}
{\cal L}&=&
\frac{f_\pi^2}{4}{\rm tr}[\partial_\mu U\partial^\mu U^\dagger] \nonumber\\
& &{} +\frac{1}{32g^2}{\rm tr}\Bigl\{[U^\dagger \partial_\mu U,U^\dagger\partial_\nu U]
[U^\dagger \partial^\mu U,U^\dagger\partial^\nu U]\Bigl\} \nonumber\\
& &{} +\frac{f_\pi^2m_\pi^2}{4}{\rm tr}[U+U^\dagger-2],
\label{lag}
\end{eqnarray}
where $ U$ is the chiral field embedding the pion field,
$f_\pi$ stands for the pion decay constant, $g$ is  dimensionless coupling constant
and $m_\pi$ denotes the pion mass.
The chiral field $ U$ can be parametrized as
\begin{eqnarray}
 U=\phi_0+i\tau_a\phi_a,
\end{eqnarray}
with $a = 1,2,3$ and the unitary relation $(\phi_0)^2+(\phi_a)^2=1$.
The pion field configuration $U$ can be rephrased in terms of quark
bilinear ones as
 \begin{eqnarray}
 \phi_0 & \sim & \bar q q, \quad
 \phi_a \sim \bar q i \gamma_5\tau_a q.
 \label{q-linear}
 \end{eqnarray}

To facilitate the later analysis, it is convenient to introduce unnormalized fields $\bar \phi_\alpha\;(\alpha=0,1,2,3)$, which are related to the corresponding normalized ones through
\begin{eqnarray}
\phi_\alpha=\frac{\bar\phi_\alpha}{\sqrt{\sum_{\beta=0}^3\bar\phi_\beta\bar\phi_\beta}}.
\end{eqnarray}

In a crystal cell, the static pion field can be expanded in terms of the Fourier series. For a crystal lattice with periodicity of $2L$ (the size of the unit cell for a single crystal), the unnormalized field $\bar{\phi}_\alpha$ goes like~\cite{Lee:2003aq}:
\begin{eqnarray}
\bar \phi_0(x,y,z)&=&\sum_{a,b,c}\bar \beta_{abc}\cos(a\pi x/L)\cos(b\pi y/L)\cos(c\pi z/L)\nonumber\\
\bar \phi_1(x,y,z)&=&\sum_{h,k,l}\bar \alpha_{hkl}^{(1)}\sin(h\pi x/L)\cos(k\pi y/L)\cos(l\pi z/L)\nonumber\\
\bar \phi_2(x,y,z)&=&\sum_{h,k,l}\bar \alpha_{hkl}^{(2)}\cos(l\pi x/L)\sin(h\pi y/L)\cos(k\pi z/L)\nonumber\\
\bar \phi_3(x,y,z)&=&\sum_{h,k,l}\bar \alpha_{hkl}^{(3)}\cos(k\pi x/L)\cos(l\pi y/L)\sin(h\pi z/L).\nonumber\\
\label{ansatz_1}
\end{eqnarray}
For a chosen crystal structure, such as the face-centered cubic (FCC) crystal in this work, the Fourier coefficient $\bar{\alpha}$ and $\bar{\beta}$ are not independent but have some
relations~\cite{Lee:2003aq}. Hereafter we shall apply these conditions to study the skyrmion crystal properties with the chiral soliton lattice.

%%%%%%%%%%%%%%%%%%%%%%%%%%%%%%%%%%%%%%%%%%%%%%%%%%%%%%%%%%%
 \subsection{CSL in skyrmion crystal}

Now we incorporate a CSL in the skyrmion crystal by introducing the neutral pion field over the skyrmion configuration $\bar U$ through the following decomposition of $U$ field
\begin{eqnarray}
U & = & \breve u\bar U \breve u ,\nonumber\\
\breve u & = & \exp\left[i\frac{\pi_3(z)\tau_3}{2}\right],
\label{fluc}
\end{eqnarray}
where $\pi_3$ is the fluctuating field of the neutral pion. To match the conventional CSL picture, we take
the $\pi_3$ as a one-dimensional configuration $\pi_3(z)$.

By substituting the chiral field $U$ with the decomposition~\eqref{fluc}
into the Skyrme Lagrangian, one has
\begin{eqnarray}
{\cal L}&=&
{\cal L}_{\rm mat}+{\cal L}_{{\rm pion}}-f_\pi^2m_\pi^2,\label{lag_w_pdw}
\end{eqnarray}
where ${\cal L}_{\rm mat}$ is the pure skyrmion matter part and ${\cal L}_{{\rm pion}}$ is
the fluctuating-pion part modified by the skyrmion matter effect, which are expressed as
\begin{eqnarray}
{\cal L}_{\rm mat} & = &
\frac{f_\pi^2}{4}{\rm tr}[\partial_\mu \bar U\partial^\mu\bar U^\dagger]\nonumber\\
& &{} +\frac{1}{32g^2}{\rm tr}\Bigl\{[\bar U^\dagger \partial_\mu\bar U,\bar U^\dagger\partial_\nu\bar U]
[\bar U^\dagger \partial^\mu\bar U,\bar U^\dagger\partial^\nu\bar U]\Bigl\}\nonumber\\
{\cal L}_{{\rm pion}} & = & {} -\frac{1}{2}A
(\partial_z \pi_3)^2+
\Bigl({f_\pi^2 m_\pi^2}B\Bigl)\cos \left(\frac{\pi_3}{f_\pi}\right)
,\label{lag_w_pdw}
\end{eqnarray}
with $A$ and $B$ being the medium modified factors from the skyrmion matter
\begin{eqnarray}
A & = & \frac{1}{3}+\left(1-\frac{1}{3}\right)(\phi_0)^2\nonumber\\
& &{} +\frac{1}{g^2f_\pi^2}
\Bigl[{} -2(\partial_x\phi_3\partial_x\phi_0+\partial_y\phi_3\partial_y\phi_0)\phi_0\phi_3 \nonumber\\
& &\qquad\qquad {} -\left\{(\partial_x\phi_0)^2+(\partial_y\phi_0)^2\right\}(\phi_0)^2\nonumber\\
& &
\qquad\qquad{}  -\left\{(\partial_x\phi_a)^2+(\partial_y\phi_a)^2\right\}(\phi_3)^2 \nonumber\\
& &\qquad\qquad{} -\sum_{i=1}^2\left\{(\partial_x\phi_i)^2 +(\partial_y\phi_i)^2\right\}(\phi_0)^2\Bigl],\nonumber\\
B&=&\phi_0.
\label{AB}
\end{eqnarray}
Generally, the ${\cal L}_{{\rm pion}}$  part includes parity odd terms such as $\sin(\pi_3/f_\pi)$ and the linear combination of $\partial_z \pi_3$. However, these terms go away when the space-average of the medium modified factors from the skyrmion matter is taken.

To study the skyrmion-matter modified properties of $\pi_3$ fields coming from the factors $A$ and $B$ in Eq.~(\ref{AB}), we use the mean field approximation. This can be achieved by taking the space averaged  defined as
\begin{eqnarray}
\langle X\rangle=
\frac{1}{(2L)^3}\int_{-L}^{L}d^3x\,X,
\end{eqnarray}
where $X$ is an arbitrary operator influenced under the static skyrmion configuration
and $\int_{-L}^L d^3x=\int_{-L}^L dx\int_{-L}^L dy\int_{-L}^L dz$. After canonically normalizing the fluctuating pion field $\pi_3$,
we find the Lagrangian for the fluctuating-pion part,
\begin{eqnarray}
\bar{\cal L}_{{\rm pion}}=-\frac{1}{2}
(\partial_z \tilde\pi_3)^2+
\Bigl({\tilde f_\pi^{*2} m_\pi^{*2}}\Bigl)\cos \left(\frac{\tilde\pi_3}{\tilde f_\pi^*}\right)
,
\label{fluc_pilag}
\end{eqnarray}
where
\begin{eqnarray}
\tilde\pi_3=\sqrt{\langle A\rangle}\pi_3,\;\;
\tilde f_\pi^{*}=\sqrt{\langle A\rangle}f_\pi,\;\;
 m_\pi^*=\sqrt{\frac{\langle B\rangle}{\langle A\rangle}}m_\pi.
 \label{mod_fpi}
\end{eqnarray}

From Lagrangian~\eqref{fluc_pilag}, one derives the equation of motion of the one-dimensional configuration $\tilde\pi_3(z)$ as
\begin{eqnarray}
\frac{\partial_z^2\tilde\pi_3}{\tilde f_\pi^*}=m_\pi^{*2}\sin\frac{\tilde\pi_3}{\tilde f_\pi^*}.
\label{eq_pion}
\end{eqnarray}
As is well known,
this equation is equivalent to the equation of a motion for a simple pendulum.
Using the Jacobi elliptic functions,
one can find
\begin{eqnarray}
-\cos\left(\frac{\tilde \pi_3}{2\tilde f_\pi^*}\right)={\rm sn}\left(\frac{m_\pi^* z}{k},k \right),
\label{config_pi}
\end{eqnarray}
where $k\,(0\leq k \leq 1)$ is the elliptic modulus.
This solution is the so-called CSL. The period of the CSL is given by
\begin{eqnarray}
l=\frac{2kK(k)}{m_\pi^*},
\end{eqnarray}
with  $K(k)$ being the complete elliptic integral of the first kind.
The CSL has the topological charge, which depends only on the boundary condition such as
$\tilde\pi_3/f_\pi^*(-l/2)=0$ and  $\tilde\pi_3/f_\pi^*(l/2) =2\pi$. The topological charge is given as
\begin{eqnarray}
Q&=&\frac{1}{2\pi}\int^{l/2}_{-l/2}dz\frac{\partial_z\tilde\pi_3}{\tilde f_\pi^*} = {}+1.
\end{eqnarray}
Hence, the CSL is the topological solution. At $k=1$, this solution corresponds to a single domain wall
which is called the pion-domain wall
as discussed in \cite{Hatsuda:1986nq,Son:2007ny}.
In this case, the period of the CSL  goes to infinity and the solution
becomes the familiar form,
\begin{eqnarray}
\tilde\pi_3 = 4f_\pi^*\arctan[\exp(m_\pi z)].
\end{eqnarray}

The energy of the CSL in the volume size $(l)^3$
 is given as
\begin{eqnarray}
E_{\rm CSL}
&=&
-\int_{-l/2}^{l/2} d^3 x \bar{\cal L}_{\rm pion} \nonumber\\
&=&
\Biggl[8\frac{f_\pi^{*2}m_\pi^*}{k} (l)^2E(k)
 +m_\pi^{*2}f_\pi^{*2}\left(1-\frac{2}{k^2}\right)(l)^3 \Biggl],
 \nonumber\\
\end{eqnarray}
where
$E(k)$ is the complete elliptic integral of the second kind.
Note that the energy of the CSL has been defined in the volume size of $(l)^3$,
 not the the skyrmion crystal size of $(2L)^3$.
This is generic because the period of the CSL ($l$) does not necessarily
coincide with the skyrmon-crystal size $2L$. Thus, to extract the contribution from the CSL to
the energy of  a single skyrmion crystal, the total energy of the skyrmoin crystal having the CSL configuration
is described as
%\begin{widetext}
\begin{eqnarray}
E_{\rm tot} & = &{} -\int_{-\infty}^\infty d^3 x {\cal L}
= N^3\left(-\int_{-L}^L d^3 x
 {\cal L}_{\rm mat}^{\rm(CSL) }\right)
,
\label{prescription}
\end{eqnarray}
%\end{widetext}
where
\begin{eqnarray}
{\cal L}_{\rm mat}^{\rm(CSL) } & = &
\frac{f_\pi^2}{4}{\rm tr}[\partial_\mu \bar U\partial^\mu \bar U^\dagger] \nonumber\\
& &{} +\frac{1}{32g^2}{\rm tr}\Bigl\{[\bar U^\dagger \partial_\mu \bar U,\bar U^\dagger\partial_\nu \bar U]
[\bar U^\dagger \partial^\mu \bar U,\bar U^\dagger\partial^\nu \bar U]\Bigl\} \nonumber\\
& &{} - a\frac{f_\pi^2m_\pi^2}{4}{\rm tr}[\bar U+\bar U^\dagger] -f_\pi^2m_\pi^2,
\label{mod_lag}
\end{eqnarray}
with
$N=\frac{\infty}{2L}$ and $a=\left[4\frac{E(k)}{k^2K(k)}
 +\left(1-\frac{2}{k^2}\right)\right]$ (for a derivation of Eq.~(\ref{prescription}), see Appendix A).
Note that in Eq.~(\ref{mod_lag}), the parameter $a$ satisfying $ a\geq {}-1$ represents the remnant of the CSL.

The per-baryon energy $E/N_B$ having the CSL structure is evaluated as
\begin{eqnarray}
E/N_B & = & \frac{1}{4}\left[E_{\rm mat} +E_{\rm pdw} + (2L)^3m_\pi^2f_\pi^2\right],\nonumber\\
E_{\rm mat} & = & {} -\int_{-L}^L d^3 x {\cal L}_{\rm mat}, \nonumber\\
E_{\rm CSL} & = & {} a\,f_\pi^{2}m_\pi^{2}
 \int_{-L}^Ld^3x\phi_0
 ,
 \label{tot_energy}
\end{eqnarray}
where $E_{\rm mat}$ stands for the energy of a single skyrmion crystal
and $E_{\rm CSL}$ denotes the energy of the CSL in a single skyrmion crystal.

As was discussed in \cite{Brauner:2016pko}, the CSL configuration spontaneously breaks the continuous translational symmetry in the z-direction as well as the continuous rotation symmetry around the z-axis, and then
the gapless mode such as phonons would appear.
Referring to the emergence of the gapless mode,
we may introduce the dynamical pion field in the skymion crystal having the CSL configuration.
From Eq.(\ref{mod_lag}), the Lagrangian incorporating the dynamical pion field under the skyrmion crystal
as the CSL reads
%~\footnote{\cmh{YLM: I should think more about this equation and relevant discussion in this paragraph.}}
\begin{eqnarray}
{\cal L}^{\rm (CSL)}&=&
\frac{f_\pi^2}{4}{\rm tr}[\partial_\mu U\partial^\mu U^\dagger]\nonumber\\
& &{} +\frac{1}{32g^2}{\rm tr}\Bigl\{[U^\dagger \partial_\mu U,U^\dagger\partial_\nu U]
[U^\dagger \partial^\mu U,U^\dagger\partial^\nu U]\Bigl\} \nonumber\\
& &{} - a\frac{f_\pi^2m_\pi^2}{4}{\rm tr}[U+U^\dagger]
-f_\pi^2 m_\pi^2,
\label{lag}
\end{eqnarray}
where the dynamical pion field has been incorporated in the chiral field $U$ as
\begin{eqnarray}
U&=&\breve u^\prime \bar U^\prime \breve u^\prime , \nonumber\\
\breve u^\prime &=&e^{i\breve\pi^a\tau^a/(2f_\pi)},
\end{eqnarray}
with $\bar U'$ being the static skyrmion configuration in the ${\cal L}^{\rm (CSL)}$
and $\breve \pi^a$ the dynamical pion field.
In Eq.(\ref{lag}), the CSL effect (specific to the z-direction)
has been rephrased as the parameter $a$.
Now one can easily see that in the derived Lagrangian,
both the translational symmetry  in the z-direction and
the continuous rotation symmetry around the z-axis
are intact even if the CSL is surely present in the skyrmion crystal.
Therefore, the gapless mode such as phonons, as mentioned above,
is actually invisible in the present approach, which is due to taking the space averaged.
%due to taking the prescription for space averaged as was done in Eq. (\ref{prescription}).
% as will clearly be presented in Appendix A.

%%%%%%%%%%%%%%%%%%%%%%%%%%%%%%%%%%%%%%%%%%%%%%%%%%%%%%%%%%%%%%%%%%%%%%%%%%%
\subsection{Baryon number density}

In the presence of the CSL, the baryon number density
\begin{eqnarray}
\rho_{B} & = &
\frac{1}{24\pi^2}\epsilon^{0\nu\rho\sigma}{\rm tr}
\Biggl[
(\partial_\nu U\cdot U^\dagger)(\partial_\rho U\cdot U^\dagger)(\partial_\sigma U\cdot U^\dagger)
\Biggl],
\nonumber\\
\end{eqnarray}
gets certainly affected.
Substituting the chiral field in Eq.(\ref{fluc}) into $\rho_B$,
we find  the modified baryon number density  as
\begin{eqnarray}
\rho_{B} & = & \rho_W
+\rho_{\rm ind},
\label{dencity_baryon}
\end{eqnarray}
where
\begin{eqnarray}
\rho_W & = & \frac{1}{24\pi^2}\epsilon^{0\nu\rho\sigma}{\rm tr}
\Biggl[
(\partial_\nu\bar U\cdot \bar U^\dagger)(\partial_\rho\bar U\cdot\bar U^\dagger)(\partial_\sigma\bar U\cdot\bar U^\dagger)
\Biggl],\nonumber\\
\rho_{\rm ind} & = & \frac{1}{2\pi^2}\frac{\partial_z\tilde\pi(z)}{f_\pi^*}\Bigl(\partial_x\phi_1\partial_y\phi_2-\partial_y\phi_1\partial_x\phi_2\Bigl),
\end{eqnarray}
with $\rho_W$ being the topological density
corresponding to the winding number and $ \rho_{\rm ind}$ being the induced baryon number density caused by the CSL.
The structure deformation of the skyrmion crystal happens
due to the presence of this induced charge,
 as will be shown later.

The baryon number is obtained by performing the spacial integration: $N= \int_{-L}^{L}d^3x \rho_B$.
From the baryon number density in the presence of CSL~\eqref{dencity_baryon}, we have
\begin{eqnarray}
N= \int_{-L}^{L}d^3x \rho_B=4,
\end{eqnarray}
because
$\int_{-L}^L d^3x \tilde \rho_{\rm ind}=0 $
\footnote{
One can then derive the following identity:
\begin{eqnarray}
&&\int_{-L}^Ldx\int_{-L}^Ldy
\Bigl(\partial_x\phi_1\partial_y\phi_2-\partial_y\phi_1\partial_x\phi_2\Bigl)
\nonumber\\
&=&
\int_{-L}^Ldx
\Bigl[\partial_x\phi_1\cdot\phi_2\Bigl]_{y=-L}^{y=L}
-\int_{-L}^Ldx\int_{-L}^Ldy
(\partial_x\partial_y\phi_1)\phi_2\nonumber\\
&&
-\int_{-L}^Ldy
\Bigl[\partial_y\phi_1\cdot\phi_2\Bigl]_{x=-L}^{x=L}
+\int_{-L}^Ldx\int_{-L}^Ldy
(\partial_x\partial_y\phi_1)\phi_2\nonumber\\
&=&
\int_{-L}^Ldx
\Bigl[\partial_x\phi_1\cdot\phi_2\Bigl]_{y=-L}^{y=L}
-\int_{-L}^Ldy
\Bigl[\partial_y\phi_1\cdot\phi_2\Bigl]_{x=-L}^{x=L}\nonumber\\
&=&0,\;\;\;\Bigl(\phi_1(L,y,z)=0,\;\phi_2(x,L,z)=0\Bigl).
\nonumber
\end{eqnarray}
Thus one can find the $\int_{-L}^L d^3x \tilde \rho_{\rm ind}=0$, hence
the baryon number is certainly conserved in the presence of the CSL.
}. This clearly shows that, as expected, the total baryon number in a crystal cell is unchanged by the presence of CSL.

%%%%%%%%%%%%%%%%%%%%%%%%%%%%%%%%%%%%%%%%%%%%%%%%%%%%%%%%%%%%%%%%%%%%%%%%%%%
\section{Numerical result}

\label{num_res}

In this section, we numerically analyze the CSL dependence on the per-baryon energy $E/N_B$ and $\langle\phi_0\rangle$ which signals the topology change in the baryonic matter and the CSL configuration. Also, the deformation of the skyrmion configuration will be discussed by examining the CSL dependence on the baryon-matter density distribution.
To work on the numerical analysis, we shall apply the following typical values for the input-model parameters~\cite{Ma:2016npf},
\begin{eqnarray}
f_{\pi}=92.4\,{\rm MeV},\;\;\;\;
g=5.93,
\end{eqnarray}
and will take the parameter $a$
 as the free parameter.

%%%%%%%%%%%%%%%%%%%%%%%%%%%%%%%%%%%%%%%%%%%%%%%%%%%%%%%%%%%%%%%%%%%%%%%%%%%
\subsection{Per-baryon energy $E_{\rm tot}/B$}

The per-baryon energy, $E_{\rm tot}/{B}$, is expressed as a function of the Fourier coefficients $\bar\beta_{abc},\bar\alpha_{hkl}^{(i)}$, as seen from Eq.~(\ref{ansatz_1}), which are used as variational parameters in the numerical calculation.
For a given set of crystal size $L$ and parameter $a$, those Fourier coefficients are determined by minimizing $E_{\rm tot}/{B}$ and, therefore, for different set of $L$ and $a$ we have different Fourier coefficients. This is how the medium-modified property of CSL is induced.

\begin{figure}[t]
\begin{center}
\includegraphics[width=8.2cm]{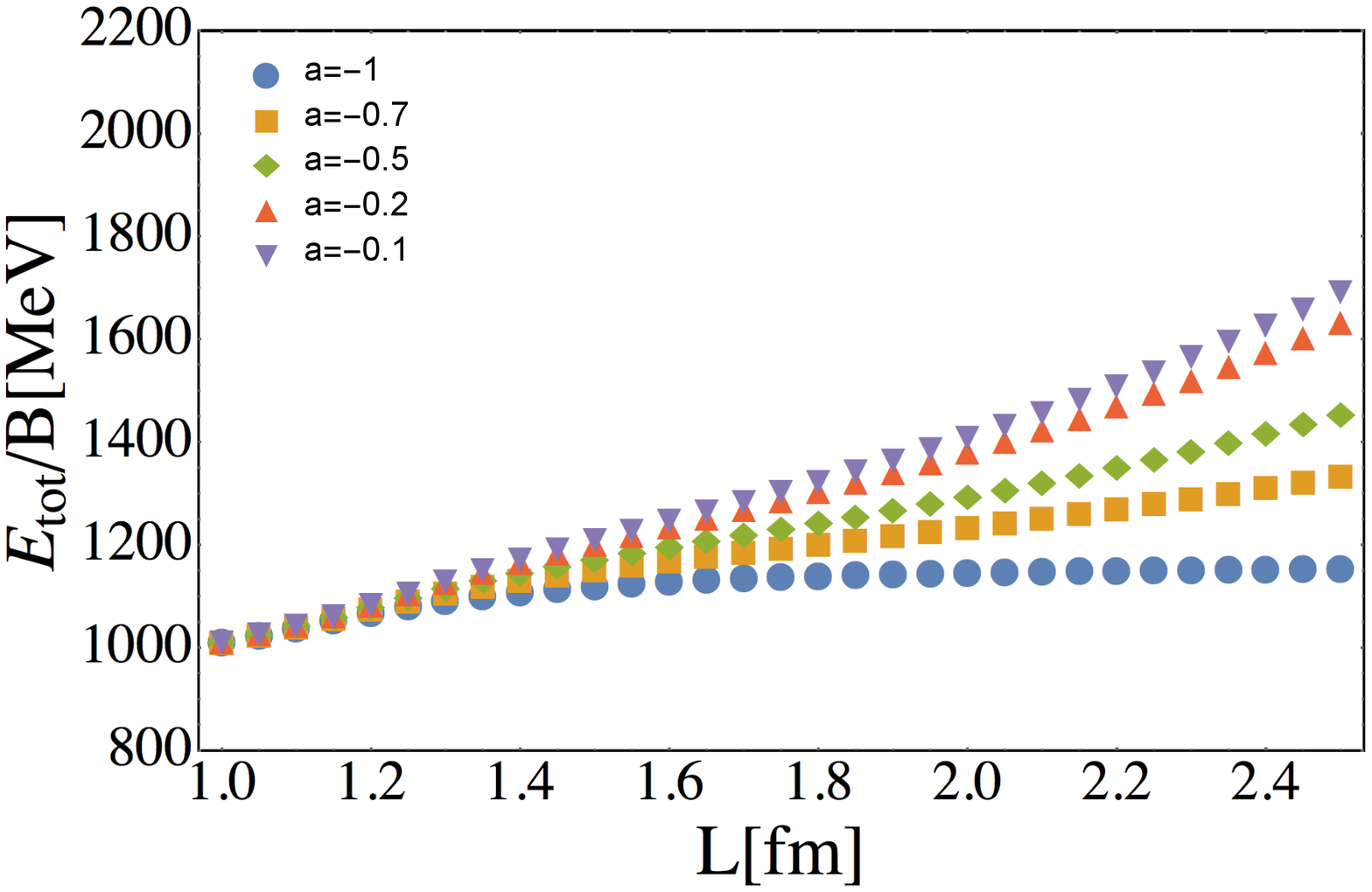}
\includegraphics[width=8.2cm]{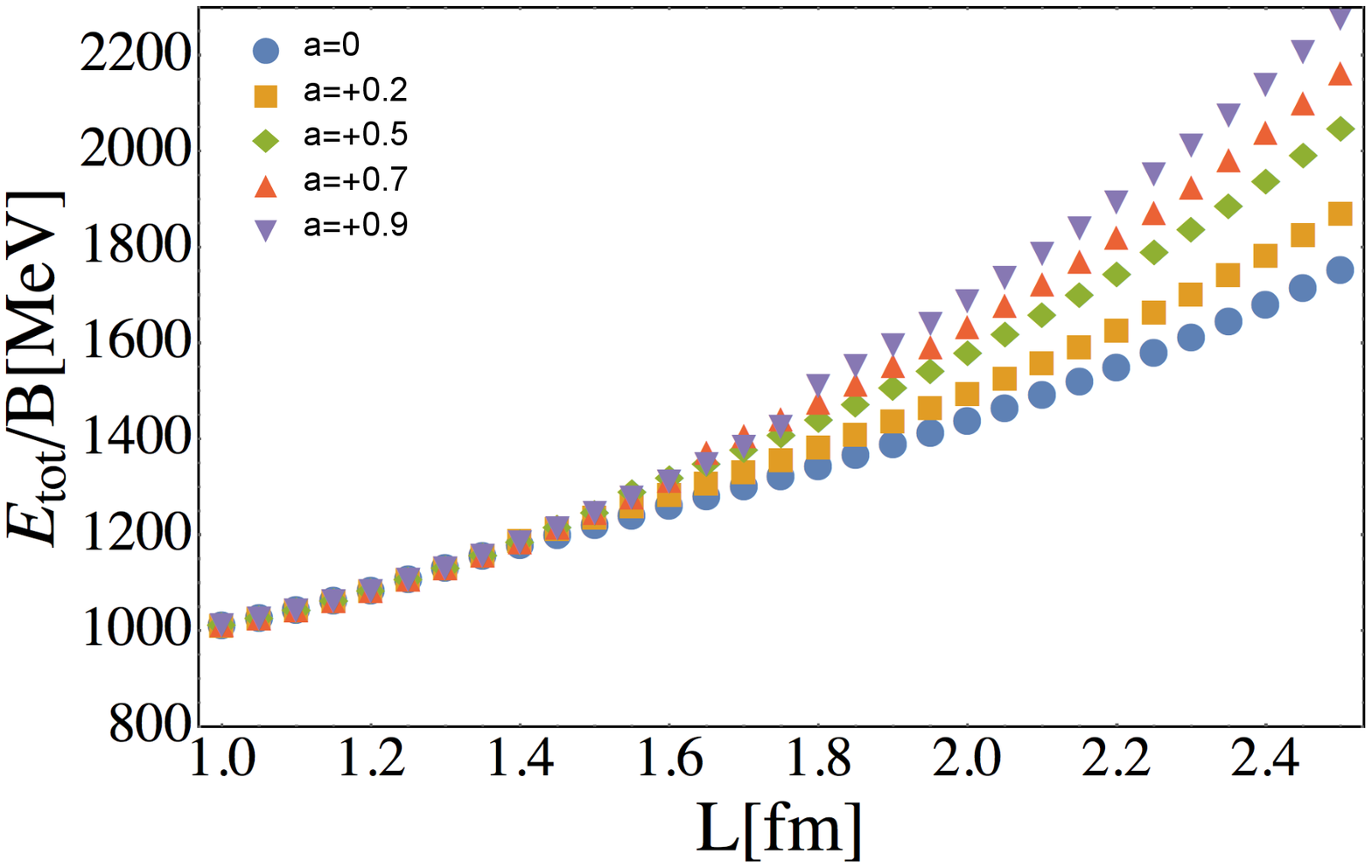}
\includegraphics[width=8.2cm]{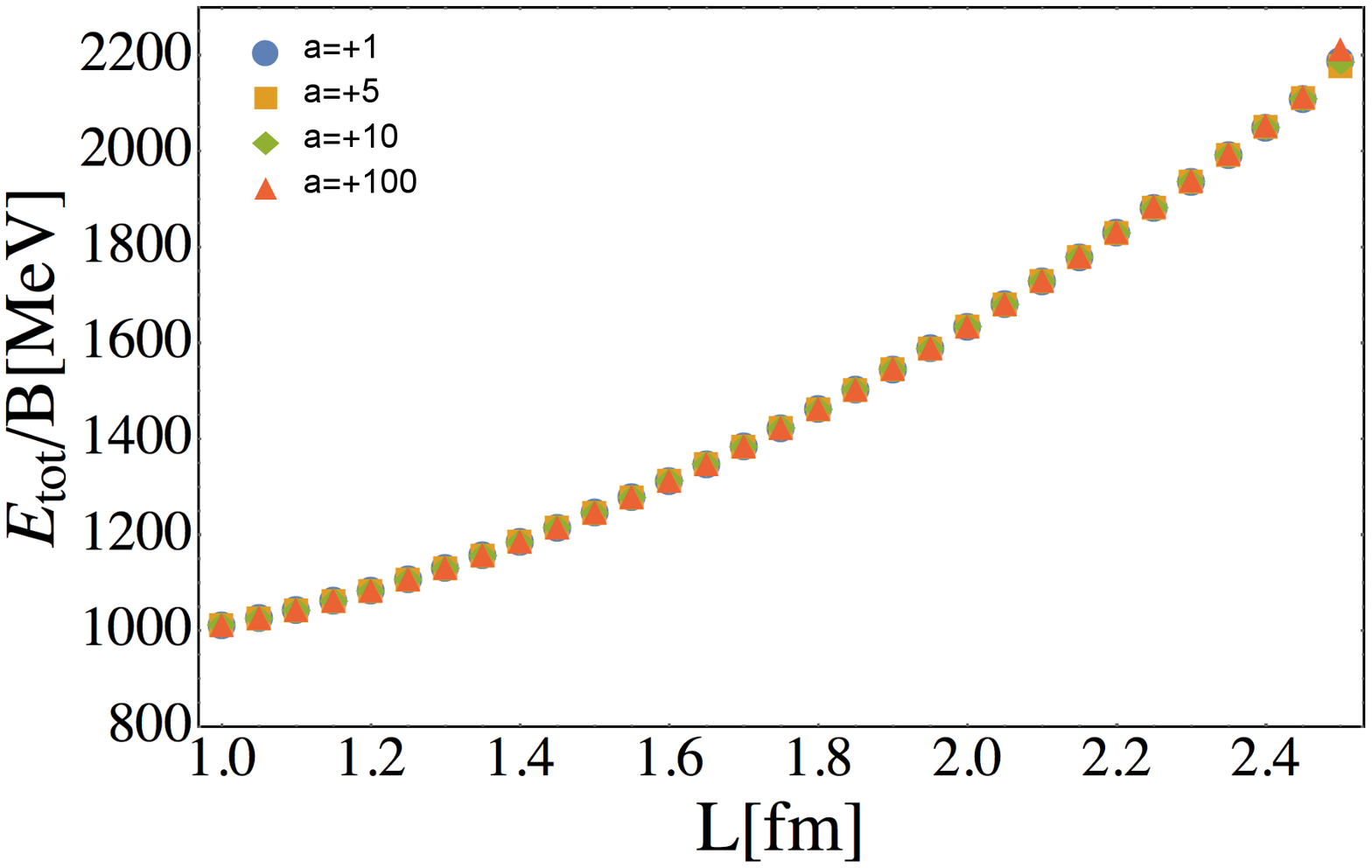}
\caption{ The per-baryon energy $E_{\rm tot}/{B}$ as a function of crystal size with ${}-1\leq a <0$ (upper panel), $0\leq a <1$ (middle panel) and $ a > 1$ (lower panel).}\label{energy_tot}
\end{center}
\end{figure}
\begin{figure}[h]
\begin{center}
\includegraphics[width=8.2cm]{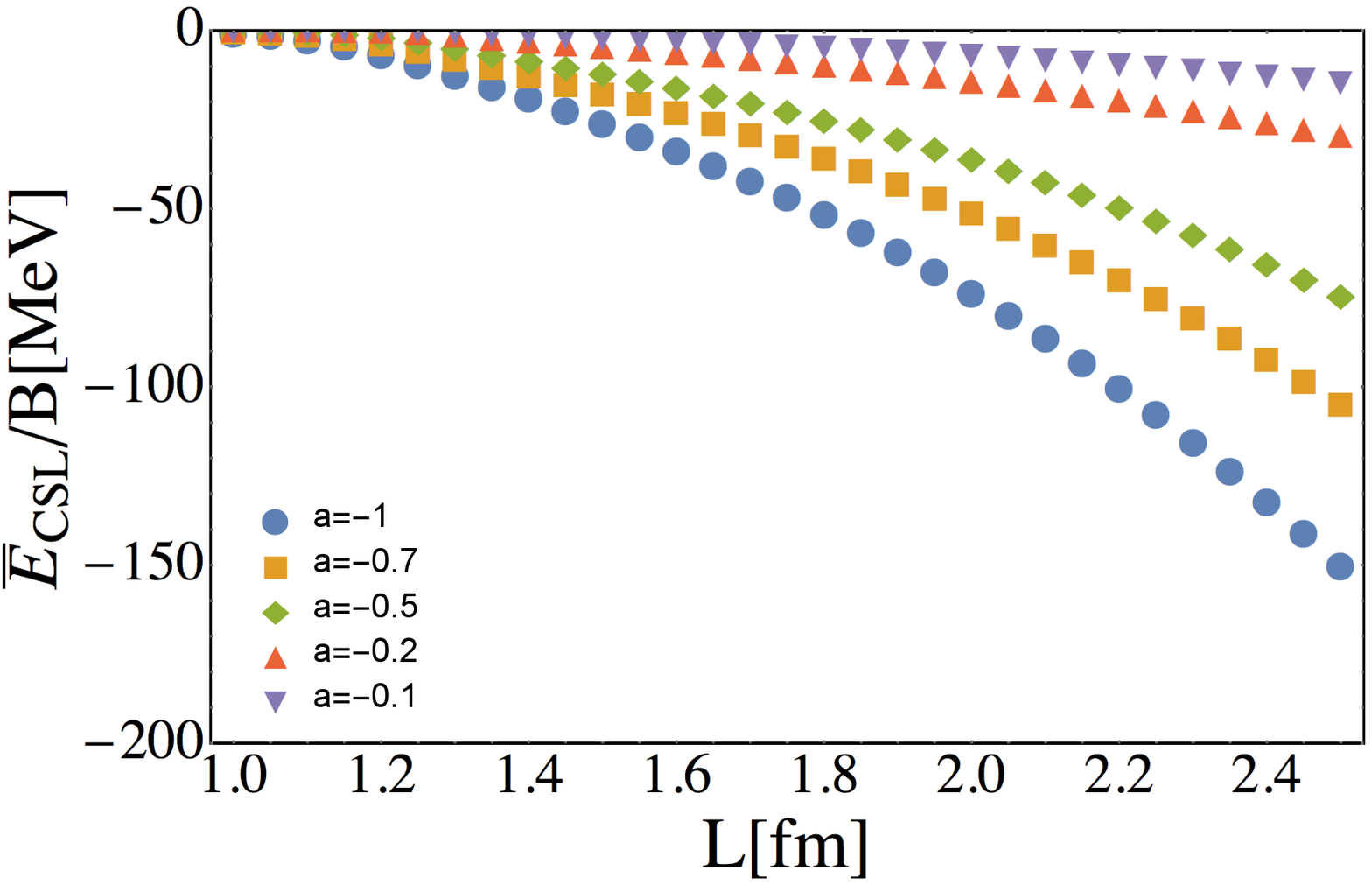}
\includegraphics[width=8.2cm]{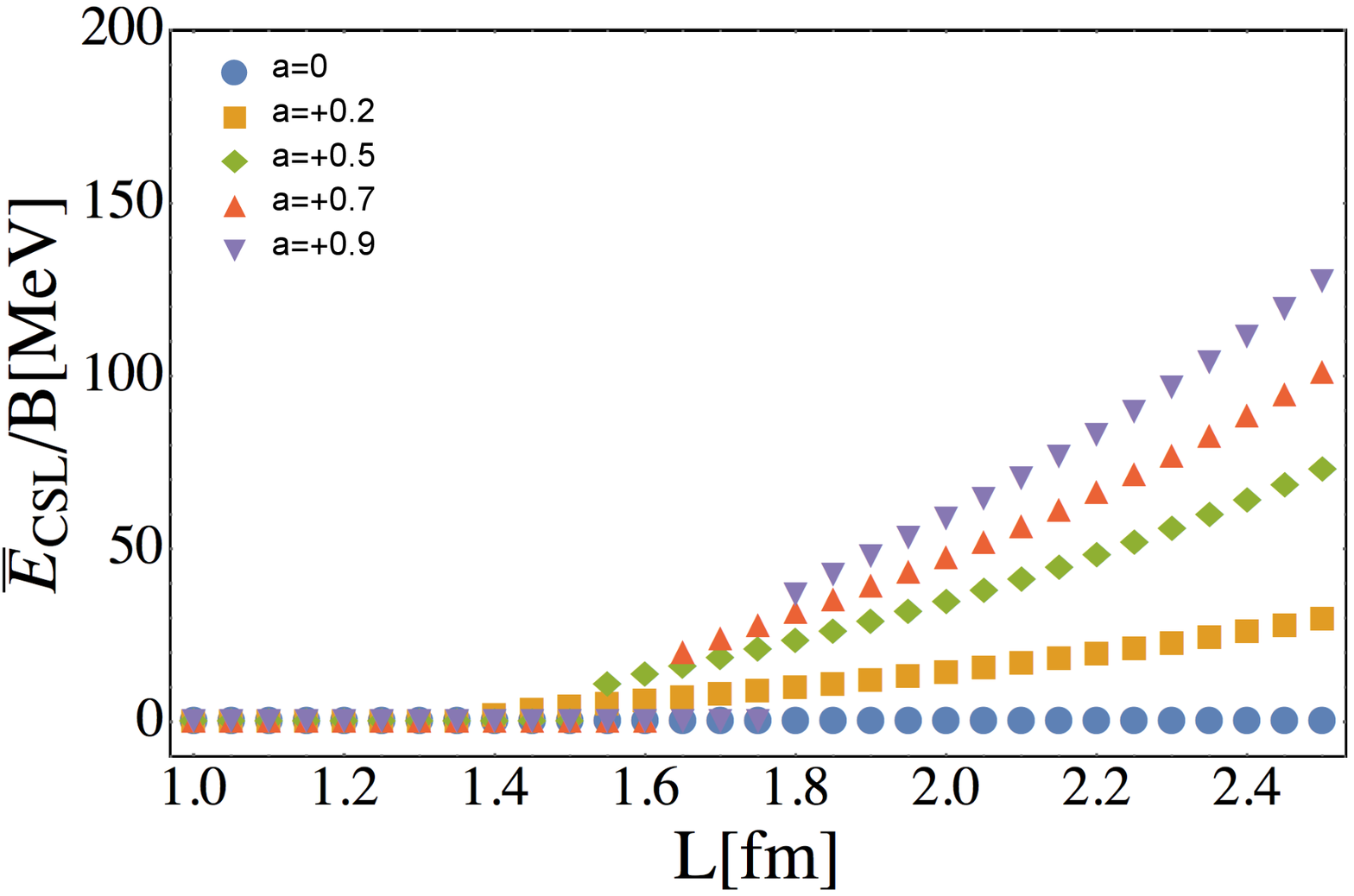}
\includegraphics[width=8.2cm]{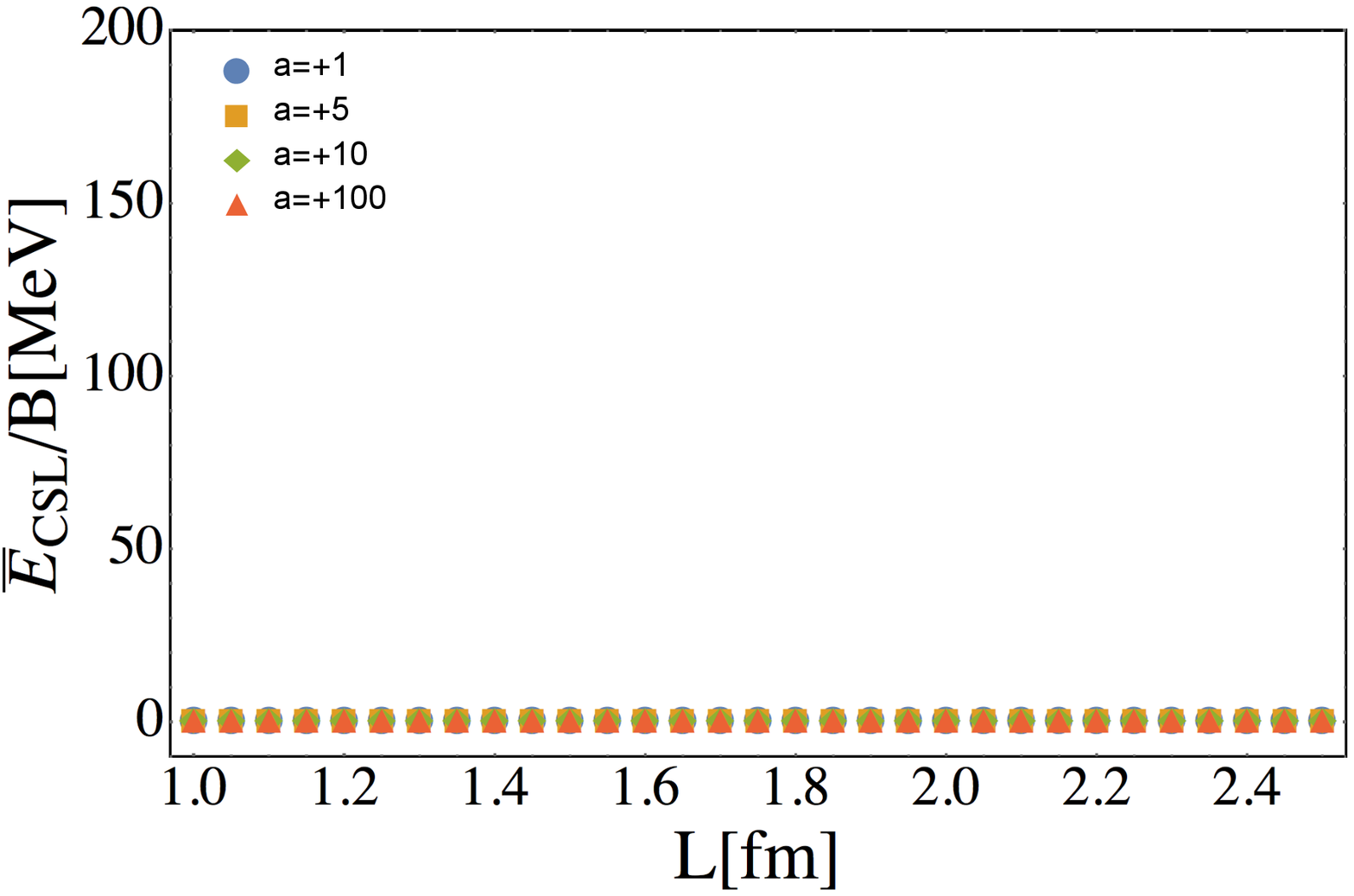}
\caption{ The per-CSL energy, $\bar E_{\rm CSL}/B$, as function of crystal size $L$ with ${}-1\leq a <0$ (upper panel), $0\leq a <1$ (middle panel) and $ a > 1$ (lower panel).}\label{CSL energy_tot}
\end{center}
\end{figure}

In Fig.~\ref{energy_tot}, we plot the per-baryon energy, $E_{\rm tot}/B$, as a function of the crystal size and the parameter $a$
\footnote{ In this study, the result obtained  in the very high density region would not be plausible, because
the $\tilde f_\pi^*$ and $m_\pi^*$ become the imaginary value,
below $L\simeq1~{\rm fm}$,
as depicted in Fig.~\ref{coef_kin}. Thus
one can not reliably
go to the high density region less than $L\simeq1~{\rm fm}$.
}.
From this figure, we find that for $a < 1$, with a fixed crystal size $L$,
 $E_{\rm tot}/{B}$ increases as a function of $a$. However, for $a \geq 1$, the value of $E_{\rm tot}/{B}$ is nearly independent of $a$ .
Note  that
this increase in $a$ can be understood by examining
the $a$-dependence of  the per-CSL energy, $\bar E_{\rm CSL}/B$ which is depicted in Fig.~\ref{CSL energy_tot}.
This phenomenon can be interpreted as follows: as will clearly been seen in Figs.~\ref{bdep25-1}--\ref{bdep12-01},
the increase of $a$ drives the enhancement of CSL wrapping with a high frequency around the skyrmion crystal,
hence  the net energy for the per-baryon gets larger as well. In addition, it is interesting
to note that although $\bar E_{\rm CSL}/B$ strongly depends on the parameter $a$ for $a < 1$ with a large $L$, it is fairly insensitive to the high density region such as $L=1.2$~fm. This feature can be understood that, at high density region, $\langle \phi_0 \rangle$ vanishes so that the CSL solution goes away.

%%%%%%%%%%%%%%%%%%%%%%%%%%%%%%%%%%%%%%%%%%%%%%%%%%%%%%%%%%%%%%%%%%%%%%%%%%%
 \subsection{Topology change on the baryonic matter}

The skyrmion crystal approach shows us a novel phenomena which is so-called
the skyrmion to half-skyrmion transition.
As the matter density increases,
the configuration of the skyrmion crystal is deformed from a face-centered cubic crystal
with one skyrmion (baryon number $1$) at each vertex
to a cubic-centered half-skyrmion crystal with a half-skyrmion (baryon-number $1/2$)~\cite{Kugler:1988mu,Kugler:1989uc}.
We will call this phenomena "topology change on the baryonic matter".
As was discussed in \cite{Lee:2003aq}, this phenomena is characterized by the space-averaged value $\langle \phi_0\rangle$: after arriving at some critical crystal size, the space-averaged value $\langle \phi_0\rangle$ vanishes,
which signals the topology change.

\begin{figure}[t]
 \begin{center}
   \includegraphics[width=8.3cm]{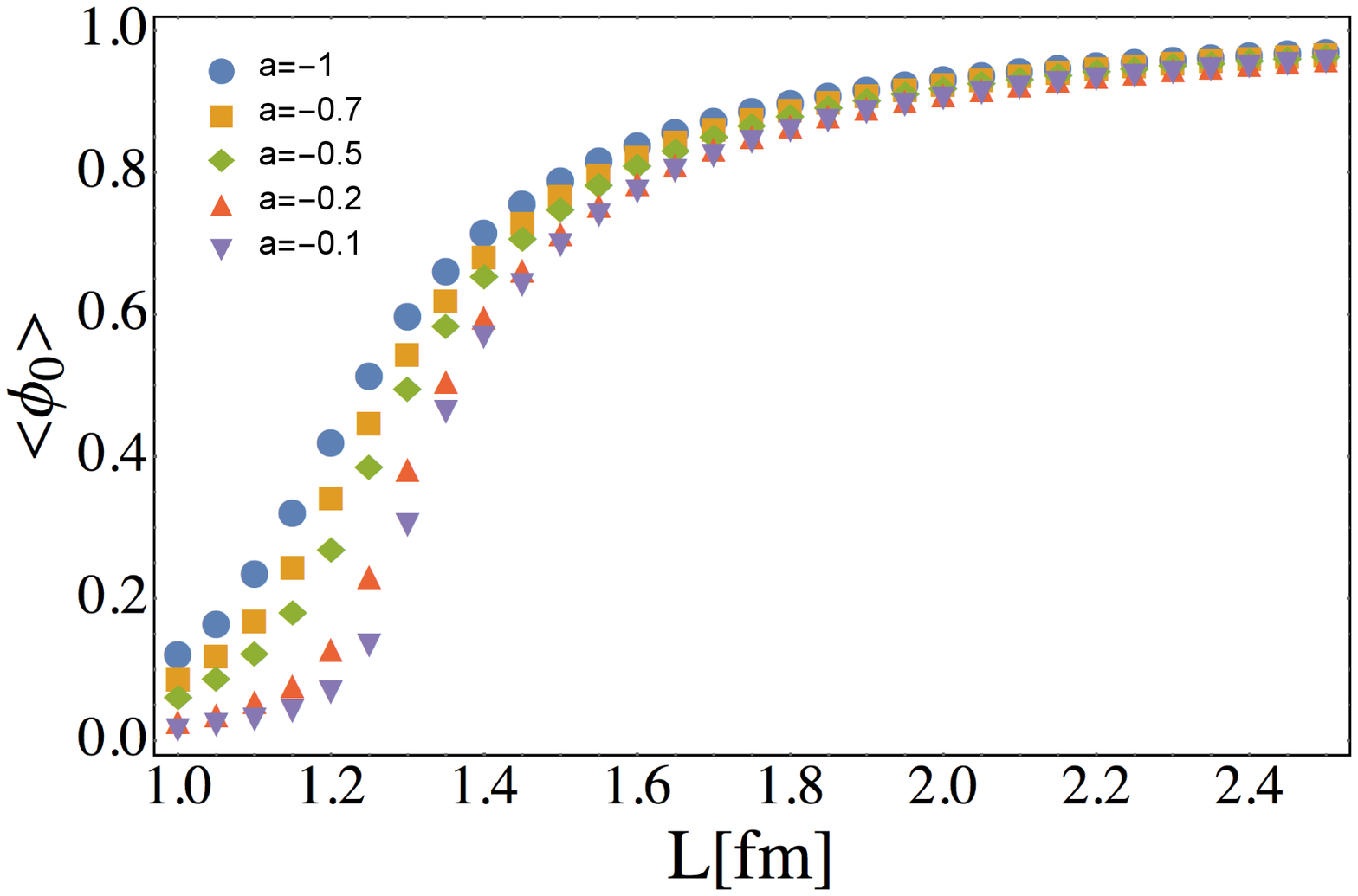}
   \includegraphics[width=8.3cm]{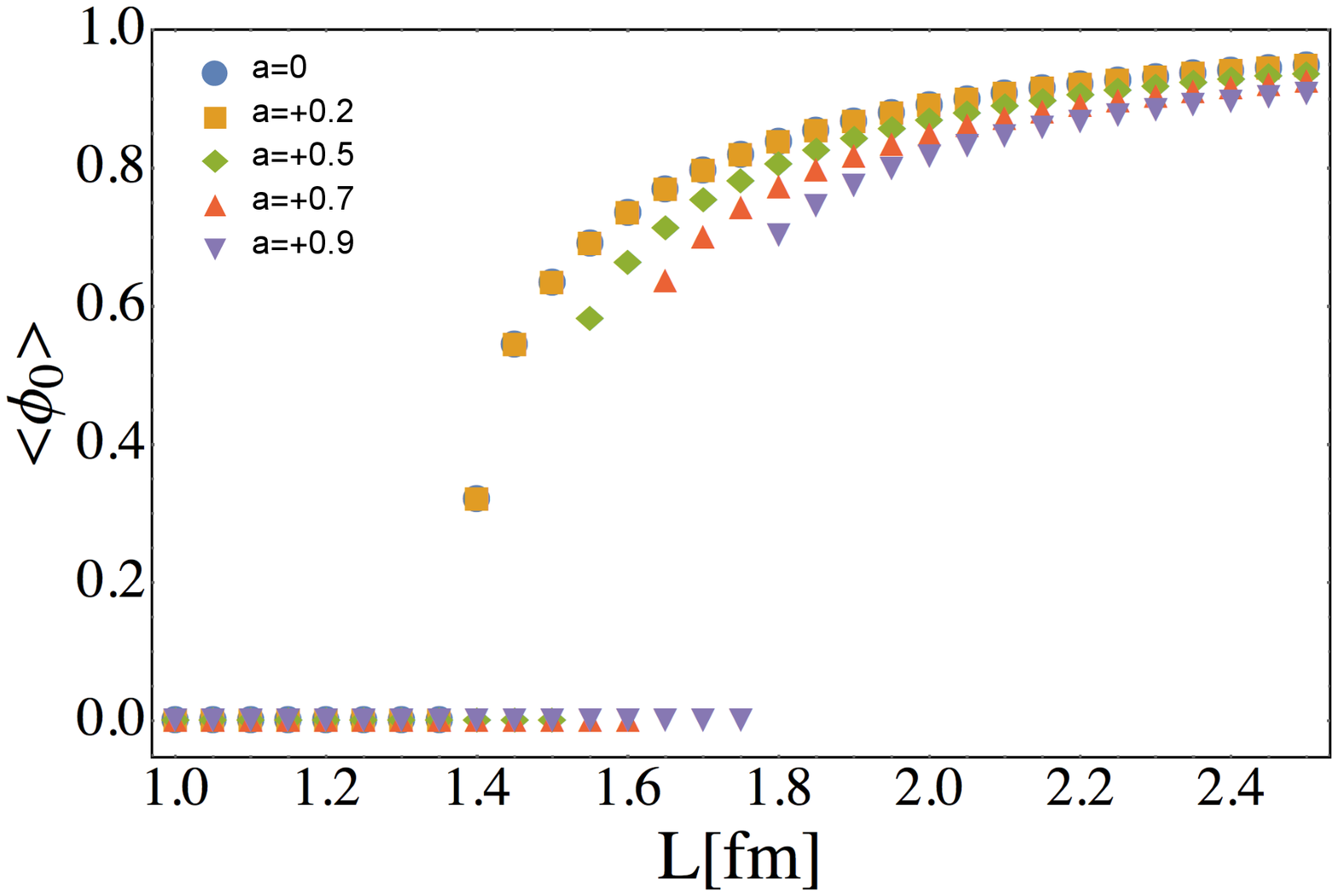}
   \includegraphics[width=8.3cm]{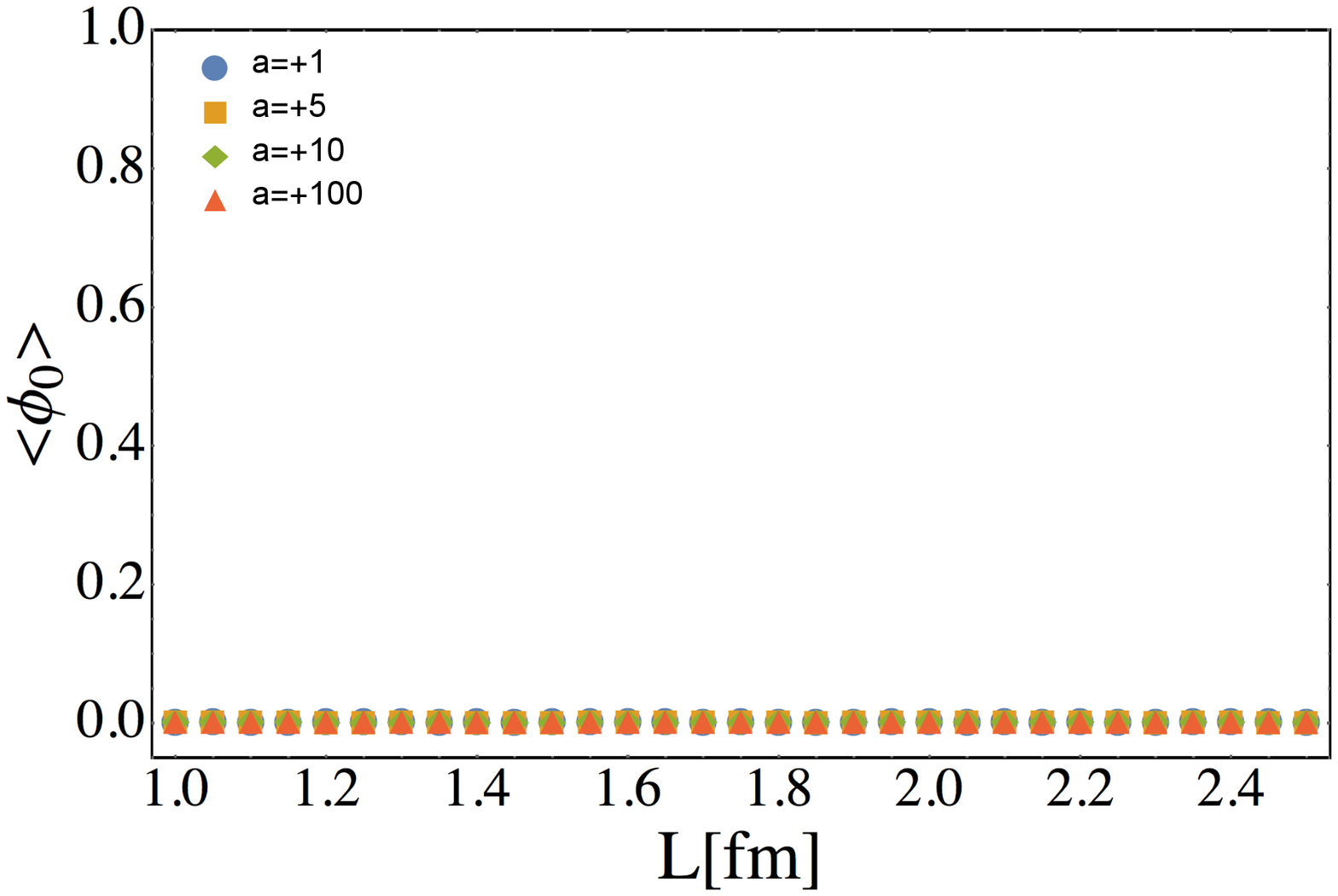}
  \end{center}
 \caption{The order parameter for topological change as a function of density with $-1\leq a <0$ (upper panel), $0\leq a <1$ (middle panel) and $1\leq a $ (lower panel).
 }
 \label{vevsigma}
\end{figure}

Fig.~\ref{vevsigma} shows the dependence of the CSL effect on the $\langle \phi_0\rangle$.
For $-1\leq a<0$, we see that as the baryon density increases, $\langle\phi_0\rangle$ asymptotically approaches to
zero. For $a=0$, the critical point of the $\langle \phi_0\rangle$ shows up at $L\simeq 1.3~{\rm fm}$,
which corresponds to the critical size $n_{1/2}\simeq 1.3 n_0$ with $n_0\simeq 0.17/{\rm fm}^3$ being
the normal nuclear matter density ($L\simeq 1.43~{\rm fm}$):
this has reproduced the result obtained in \cite{Lee:2003aq}. As the parameter $a$ increases up to $a=1$,
the topology change point is shifted to a low density region
(for instance $n_{1/2}\simeq 0.55n_0$ when $a=0.9$) and the value of the  $\langle \phi_0\rangle$
gets smaller.
This implies that the CSL causes an inverse catalysis for the topology change on
 the baryonic matter.
And, above $a=1$, the $\langle\phi_0\rangle$ vanishes for any crystal size.
This implies that
the parameter $a$ also has  the critical point for the vanishing $\langle\phi_0\rangle$,
 which is found as $a_c=1$.

 %%%%%%%%%%%%%%%%%%%%%%%%%%%%%%%%%%%%%%%%%%%%%%%%%%%%%%%%%%%%%%%%%%%%%%%%%%%
 \subsection{Profile of chiral soliton lattice  in the baryonic matter}
 \label{profile_pion}

 We next discuss the matter effect on the profile of the CSL
 given by the normalized pion field $\tilde\pi_3$ in Eq.(\ref{config_pi}), which
 is the function of the position space, specified to the z-direction,
  and the elliptic modulus $k$ (or the parameter $a$).
The CSL feels the matter effect through the medium-modified pion decay constant
$\tilde f_\pi^*$ and  the effective pion mass, $m_\pi^*$. Namely,
the CSL  implicitly depends on the crystal size, $L$.

\begin{figure}[t]
  \begin{center}
   \includegraphics[width=8.5cm]{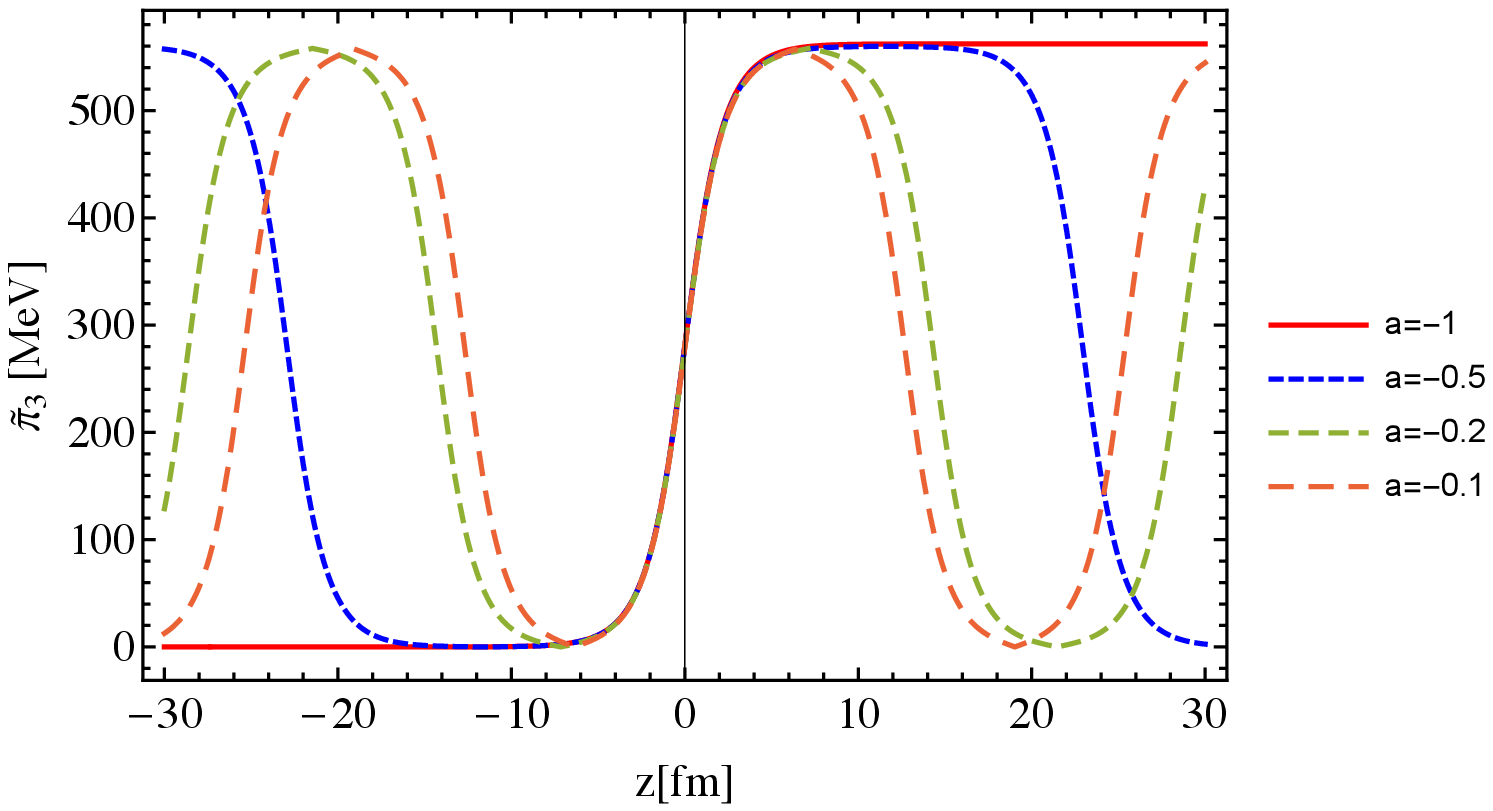}
   \includegraphics[width=8.5cm]{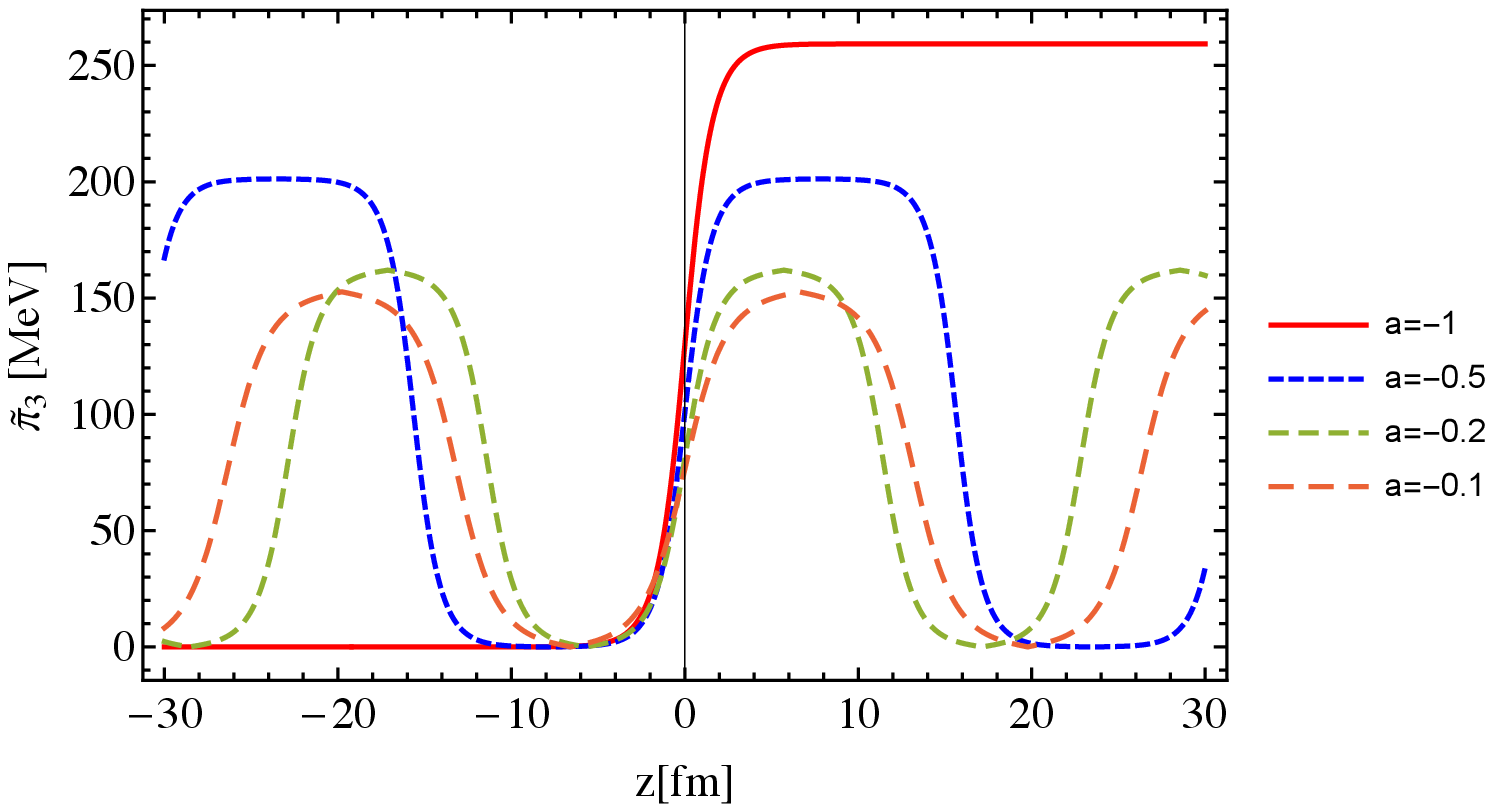}
 \end{center}
  \caption{
The profile of the $\tilde\pi_3(z)$ with $L=2.5~{\rm fm}$ (upper panel) and $L=1.2~{\rm fm}$(lower panel),
 in the region where $-1\leq a <0$.
 }
 \label{prof_pi1}
\end{figure}

\begin{figure}[t]
\begin{center}
  \includegraphics[width=8.5cm]{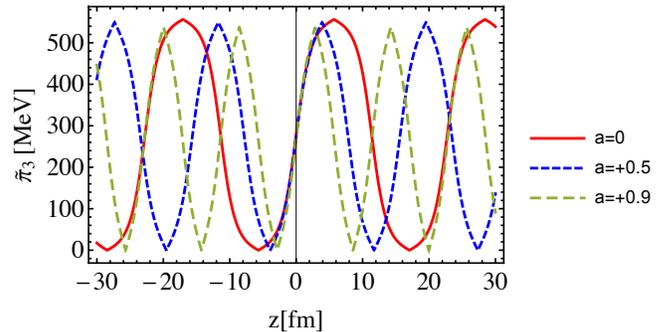}
  \end{center}
 \caption{ The profile of the $\tilde\pi_3(z)$
 at the low density region, $L=2.5~{\rm fm}$,
 in the region where $0\leq a <1$.
 }
  \label{prof_pi3}
\end{figure}

\begin{figure}[h]
  \begin{center}
   \includegraphics[width=8.5cm]{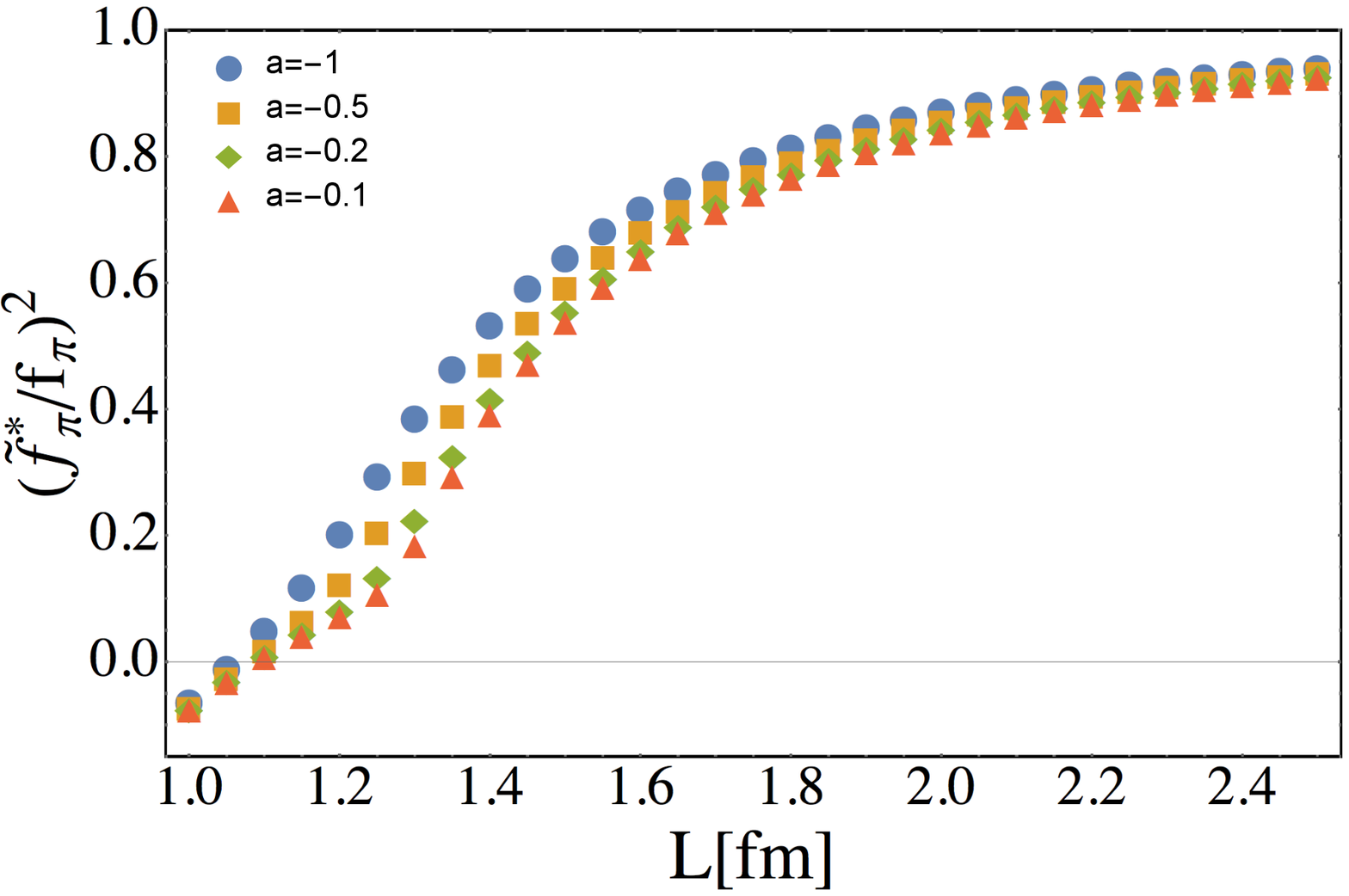}
   \includegraphics[width=8.5cm]{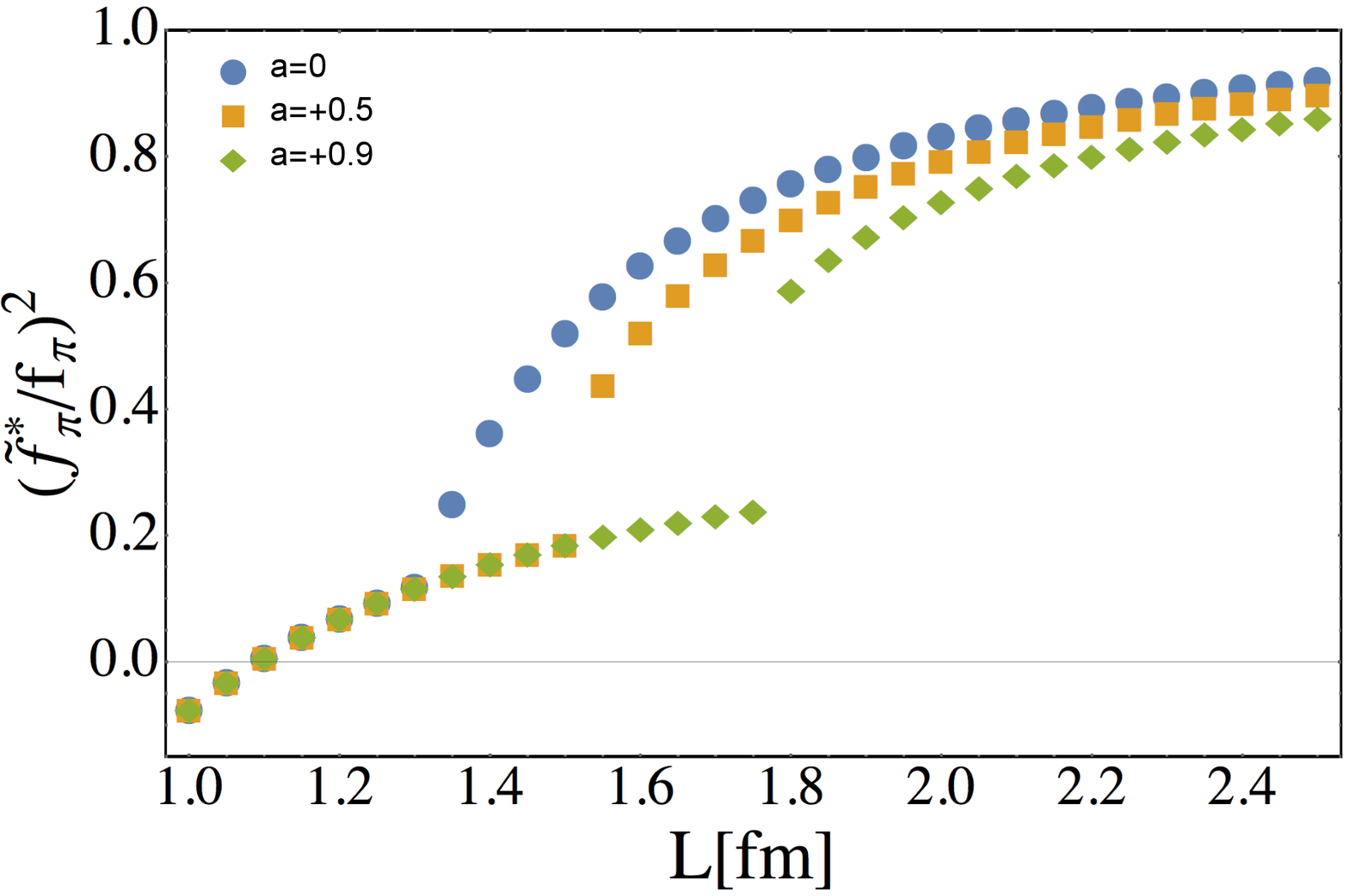}
 \end{center}
  \caption{
  The crystal size dependence of  $(\tilde f^*_\pi/f_\pi)^2$ in the region ${}-1\leq a<0$ (upper panel) and $0\leq a <1$ (lower panel).
 }
 \label{coef_kin}
\end{figure}

We plot in Figs.~\ref{prof_pi1} and~\ref{prof_pi3} the profile of the CSL
for several different values of $a$ and $L$. As the parameter $a$ increases, the period of the CSL becomes smaller for a fixed $L$. As for the amplitude of the CSL, it does not change in a low density region even if the parameter $a$ increases. In contrast, in a high density region (e.g. $L=1.2~{\rm fm}$), the amplitude becomes small as the parameter $a$ increases up to $a=0$. The suppression of the amplitude for the CSL might be somewhat correlated  with the chiral restoration. As depicted in Fig.~\ref{coef_kin}, the $\tilde f_\pi^{*2}$ normalized to the (square of) pion decay constant $f_\pi^2$, defined as in Eq.~(\ref{mod_fpi}), gets smaller than unity when the parameter $a$ becomes larger, which could be in relation to a signal of chiral restoration.

Note also that in  a high density region for $0\leq a <1$ and above the critical point $a_c=1$ for any crystal size, the CSL configuration goes away. Because the mass term in the equation of motion for the $\tilde\pi_3$ vanishes when $\langle \phi_0\rangle$ in a high density region. The pion field $\tilde\pi_3$ is not able to have the nontrivial CSL solution.

Recall that we are analyzing the properties of the skyrmion crystal assuming  existence of the non-trivial CSL. Therefore we need to discard the result obtained in a high density region for $0\leq a <1$ and above the critical point $a_c=1$ for any crystal size.

%%%%%%%%%%%%%%%%%%%%%%%%%%%%%%%%%%%%%%%%%%%%%%%%%%%%%%%%%%%%%%%%%%%%%%%%%%%
\subsection{Vacuum stability in skyrmion crystal with chiral soliton lattice }

We next discuss the validity for numerical results in terms of
the vacuum stability in skyrmion crystal with CSL.
To check the stability condition of the dynamical pion field $\breve \pi$,
we expand the Lagrangian in Eq.~(\ref{lag}) up to the second order in the $\breve \pi$,
and then the pion Lagrangian is found as
\begin{eqnarray}
{\cal L}^{({\rm pion})} & = &
\frac{1}{2}C\partial_t\breve\pi^a\partial_t\breve\pi^a
-\frac{1}{2}C^\prime\partial_i\breve\pi^a\partial_i\breve\pi^a \nonumber\\
& &{} +
a\langle\phi_0\rangle{f_\pi^2 m_\pi^2}\breve\pi^a\breve\pi^a,
\end{eqnarray}
where $C$ and $C^\prime$ are space averaged products composed
of the static skyrmion configuration $\bar U^\prime$.
One can easily find that the $C$ is positive value for any crystal size so that
the sign of the pion mass term,
$a\langle\phi_0\rangle{f_\pi^2 m_\pi^2}\breve\pi^a\breve\pi^a$,
should be minus to stabilize the dynamical pion field~\footnote{Actually if the sign of the pion mass term changes to be plus, the dynamical pion field in the
skyrmion crystal having the CSL becomes a tachyonic particle.}.
In the region where $-1<a\leq0$, the dynamical pion field is stabilized due to
$\langle\phi_0\rangle>0$ for any crystal size. However, in the region where $a>0$,
the dynamical pion becomes a tachyonic particle.
As as result, to stabilize the baryonic matter with the CSL, the parameter $a$ has to be constrained in the range $-1\leq a\leq 0$.

%%%%%%%%%%%%%%%%%%%%%%%%%%%%%%%%%%%%%%%%%%%%%%%%%%%%%%%%%%%%%%%%%%%%%%%%%%%
\subsection{Deformation of the skyrmion configuration}

We finally discuss the effect of the CSL structure evaluated in subsection~\ref{profile_pion} on the skyrmion crystal configuration and the single baryon shape in the presence of the CSL.

In the skyrmion crystal approach, the skyrmion configuration can be extracted by examining the baryon-number
density-distribution functions in Eq.~\eqref{dencity_baryon}.
From Eq.~\eqref{dencity_baryon}, one can easily check that
the winding number density  $\rho_W$ keeps the crystal symmetries
for the FCC and CC structures in presence the CSL.
On the other hand, for the induced-baryon number density
$\rho_{\rm ind}$, the crystal symmetries are explicitly broken by the CSL.
This indicates that a skyrmion configuration is deformed by the CSL configuration.

\begin{figure}[!htpb]
  \begin{center}
   \includegraphics[width=5.5cm]{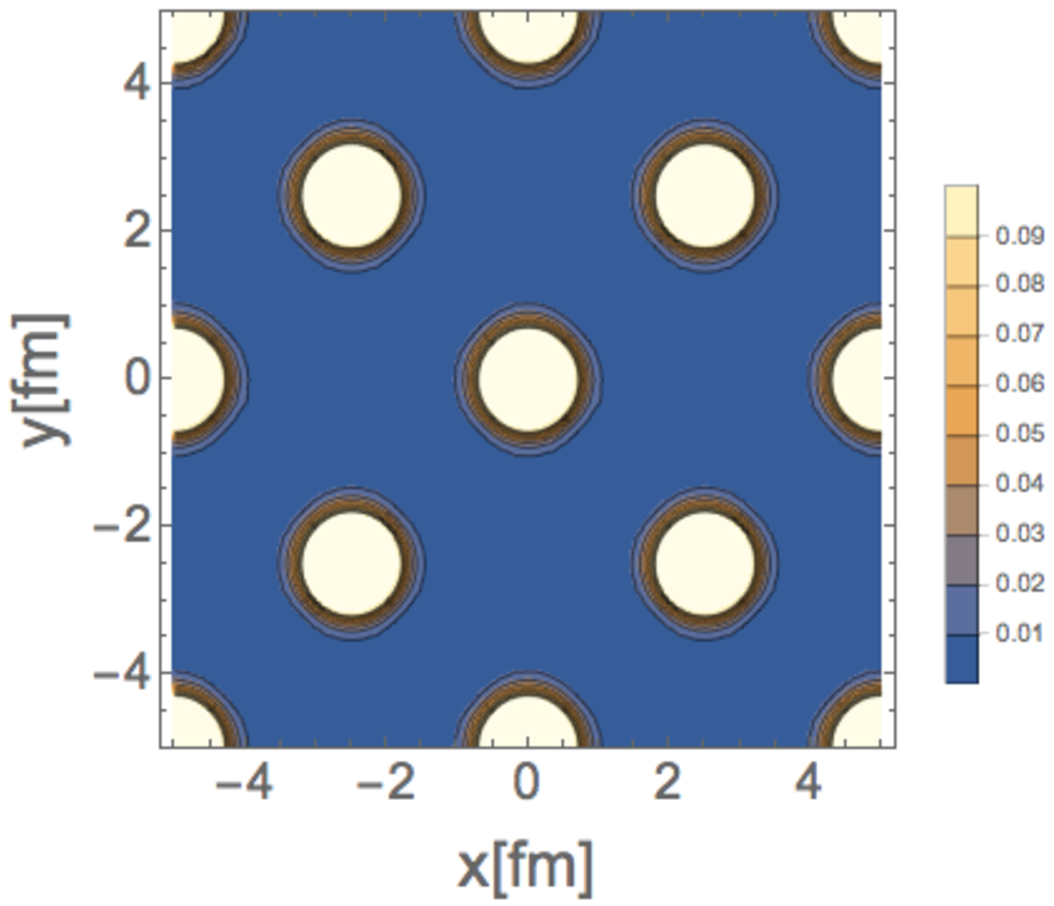}
   \includegraphics[width=5.5cm]{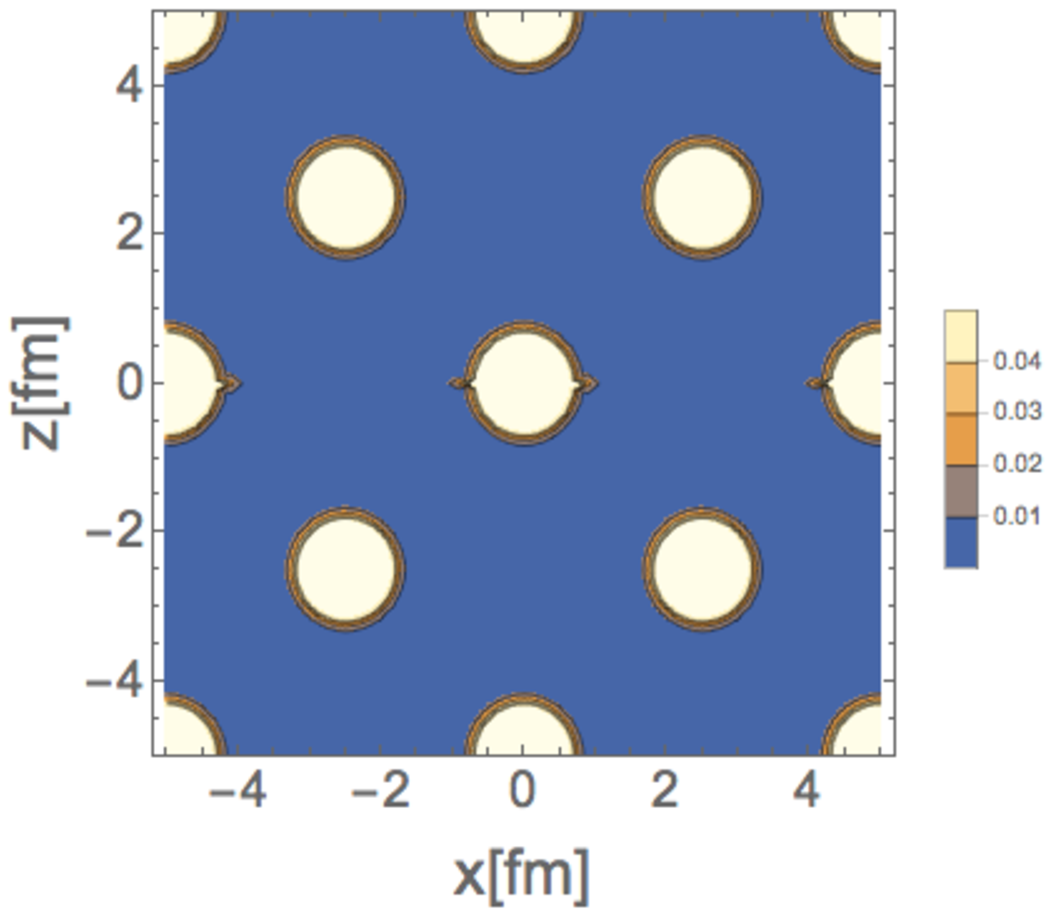}
   \includegraphics[width=5.5cm]{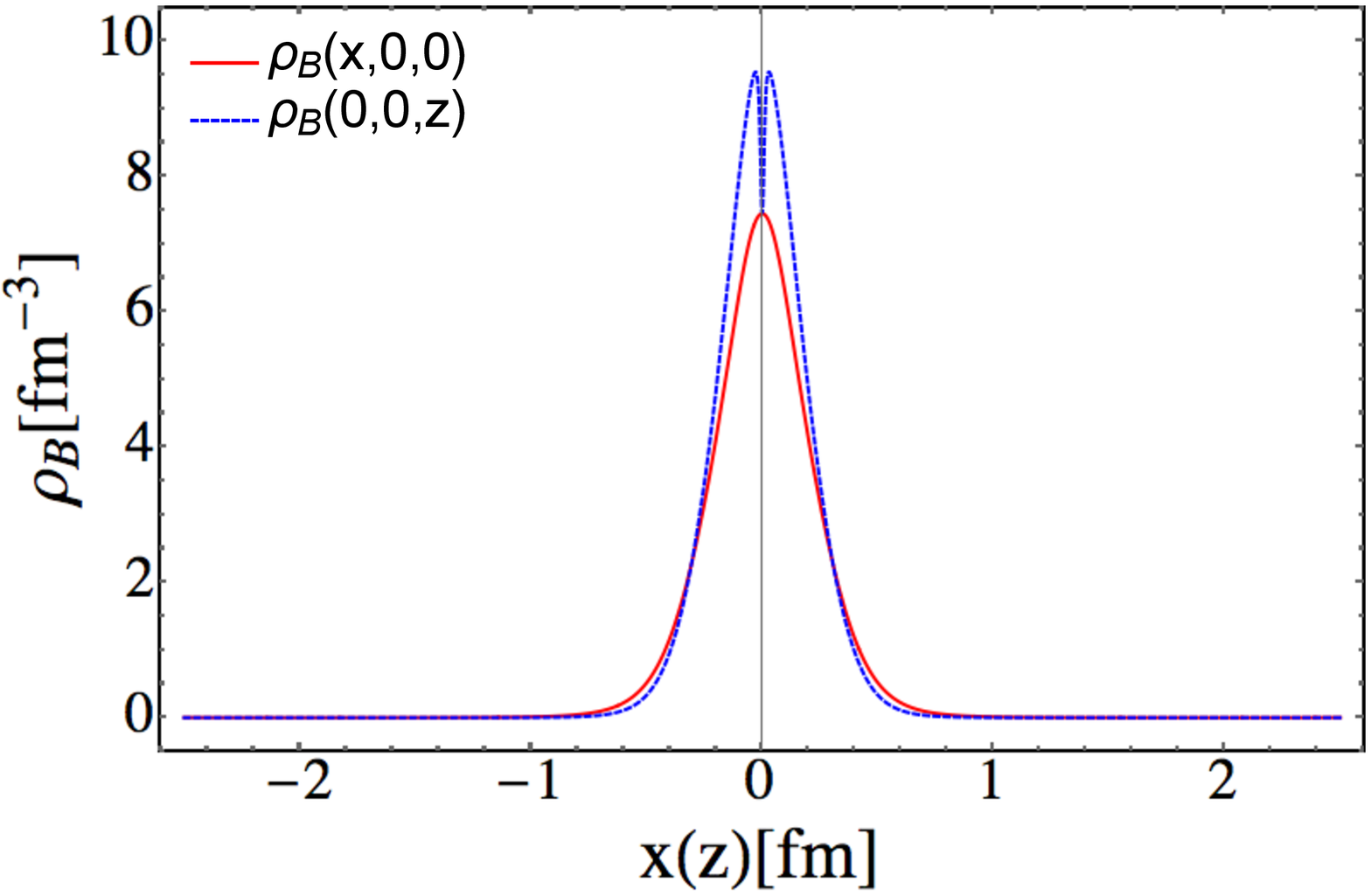}
   \end{center}
 \caption{
 The skyrmion configuration at $L=2.5~{\rm fm}$ and $a=-1$.
The upper panel displays the density contour plot on x-y plane specified at $z=0$, the middle panel is the density contour plot on the x-z plane specified at $y=0$, and the lower panel corresponds to the distribution
 of the single baryon along x-axis or z-axis.
 }
 \label{bdep25-1}
\end{figure}

\begin{figure}[!htpb]
  \begin{center}
   \includegraphics[width=5.5cm]{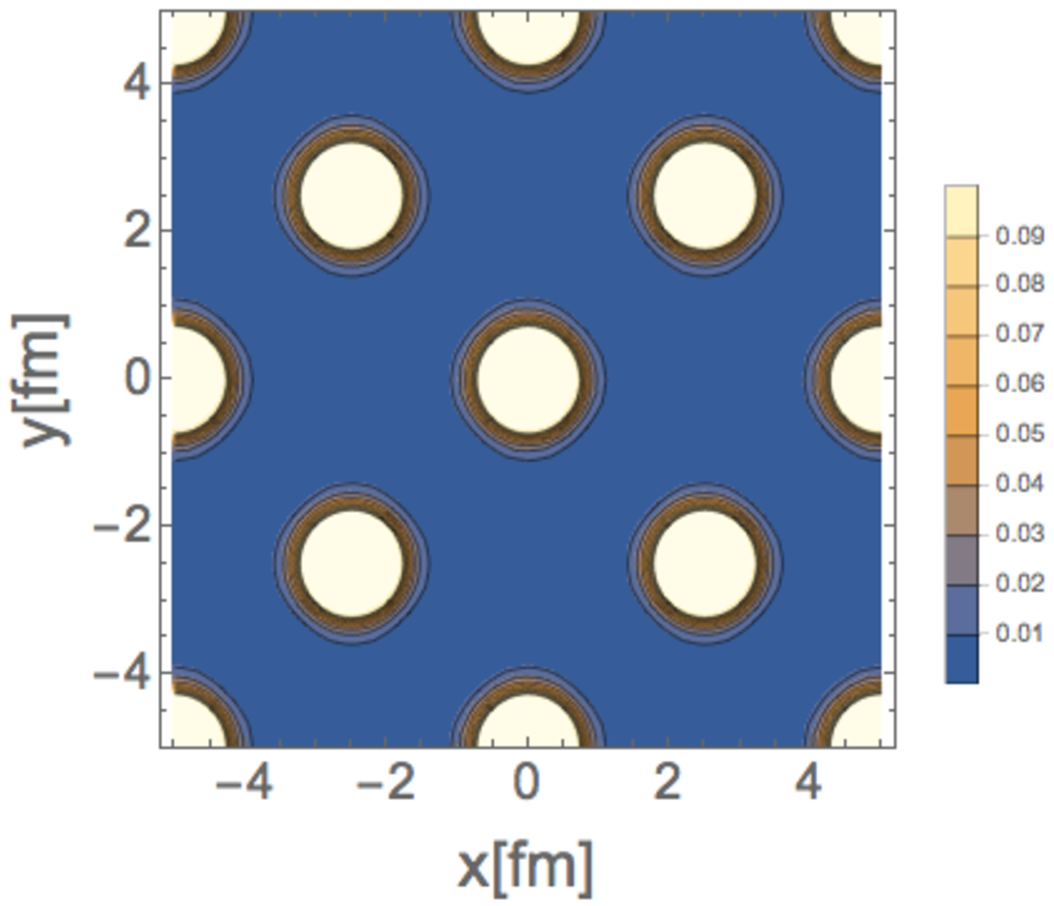}
   \includegraphics[width=5.5cm]{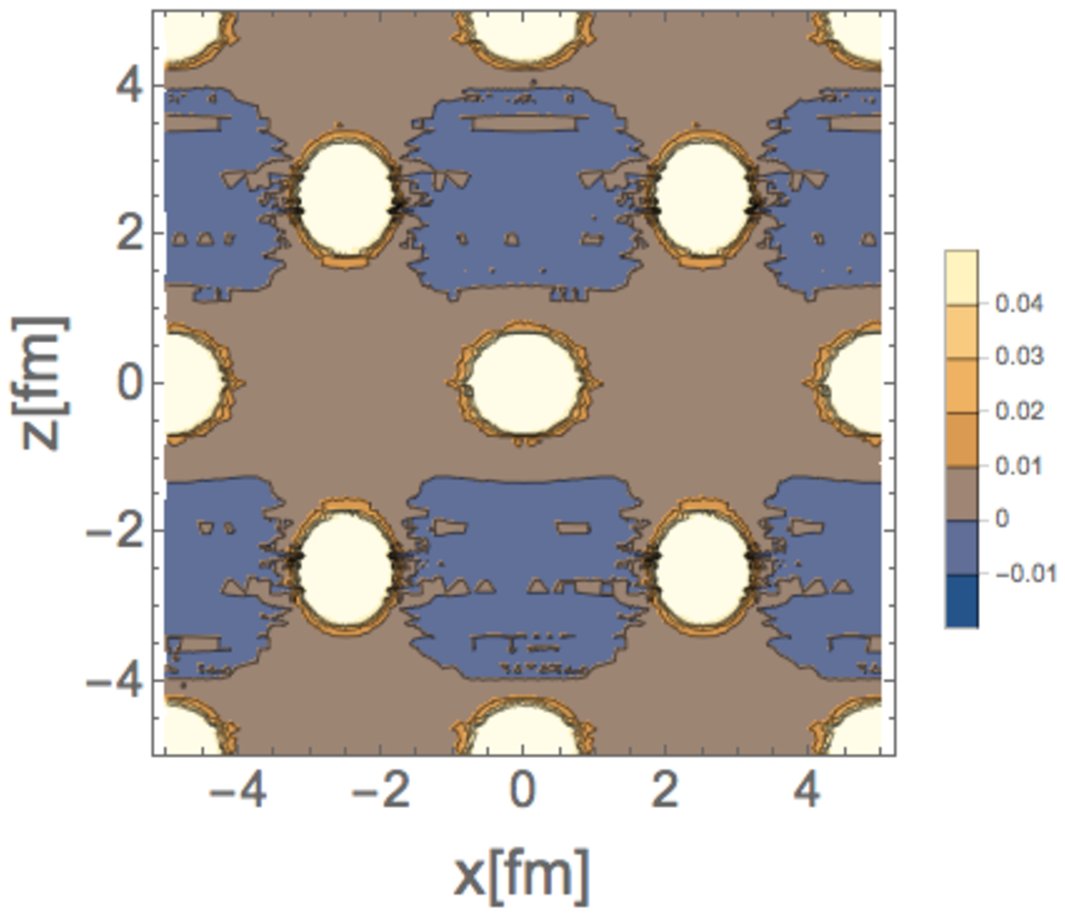}
   \includegraphics[width=5.5cm]{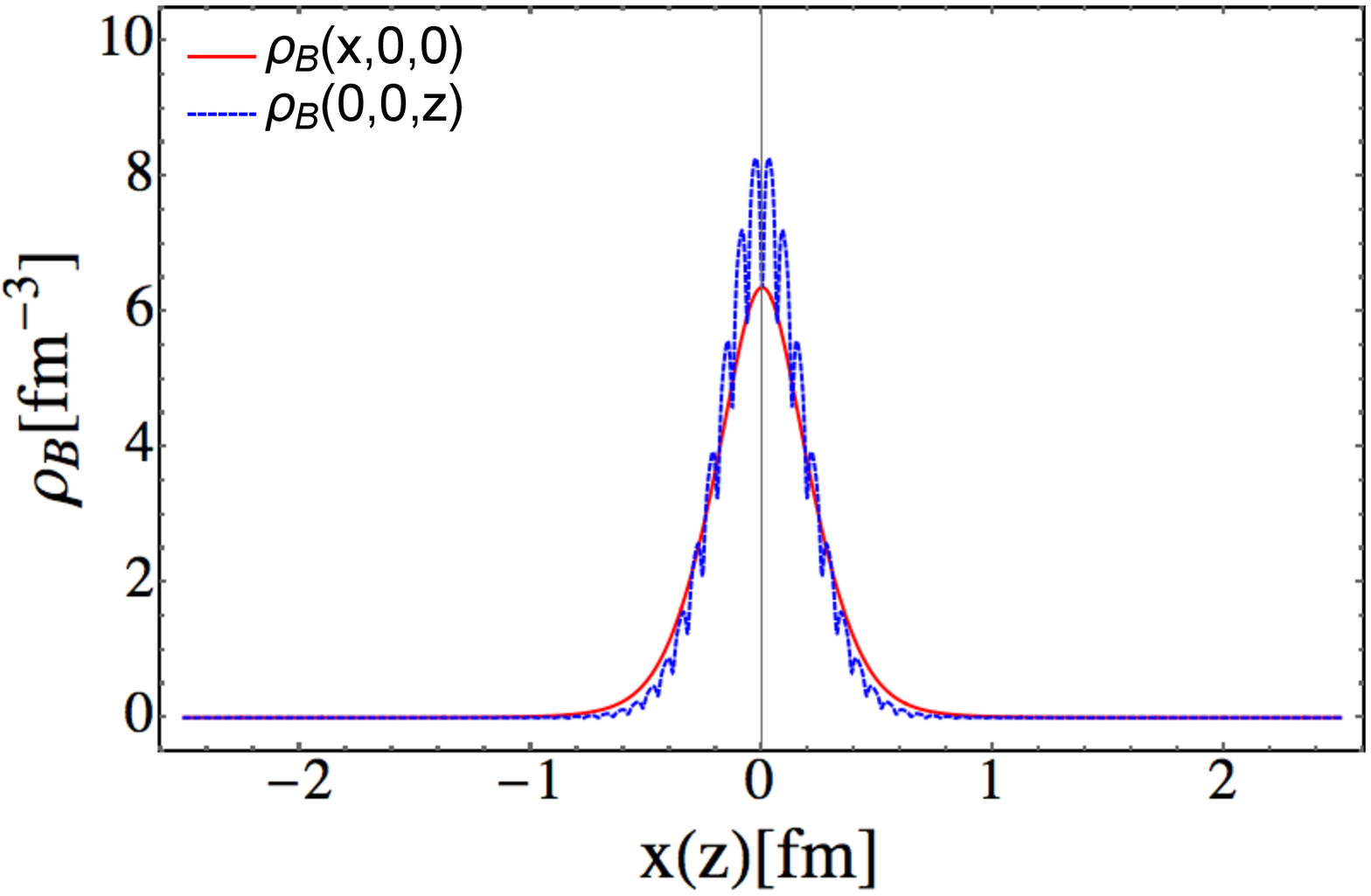}
  \end{center}
 \caption{
 The same as Fig.~\ref{bdep25-1} but with $a={}-0.1$. }
 \label{bdep25-01}
\end{figure}

\begin{figure}[!htpb]
  \begin{center}
   \includegraphics[width=5.5cm]{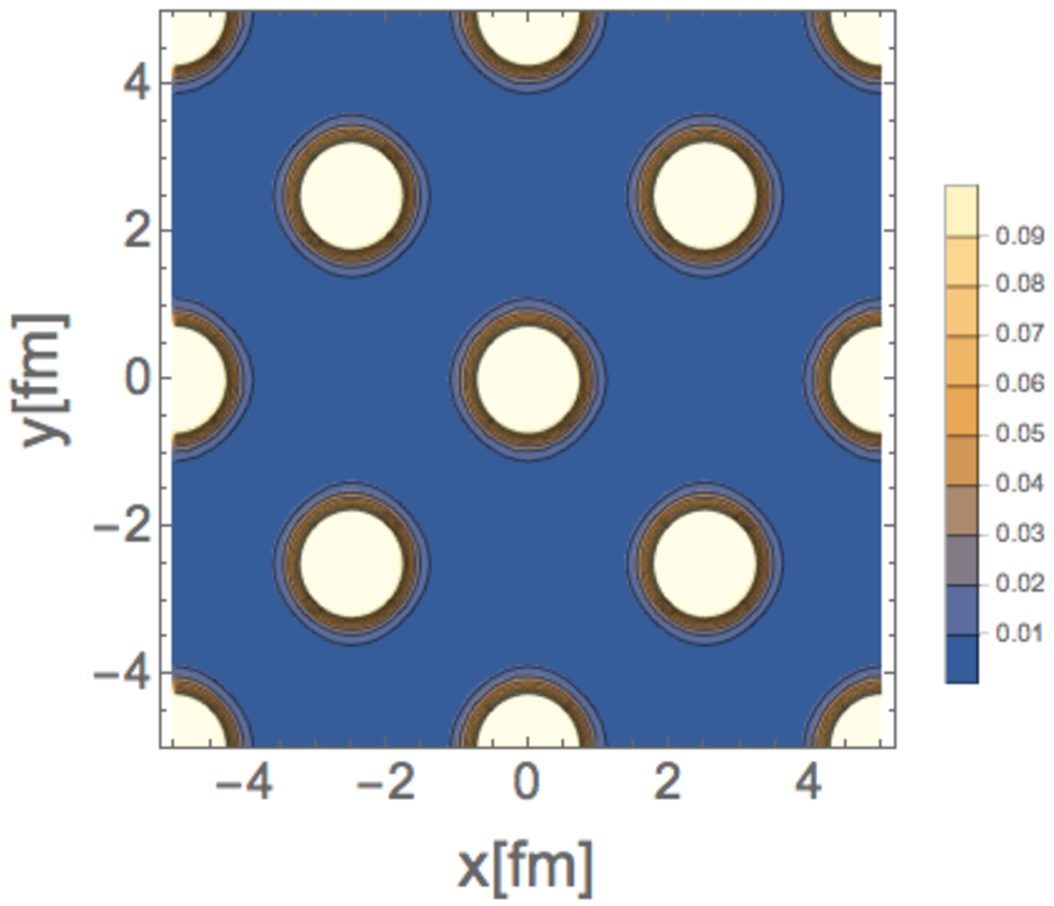}
   \includegraphics[width=5.5cm]{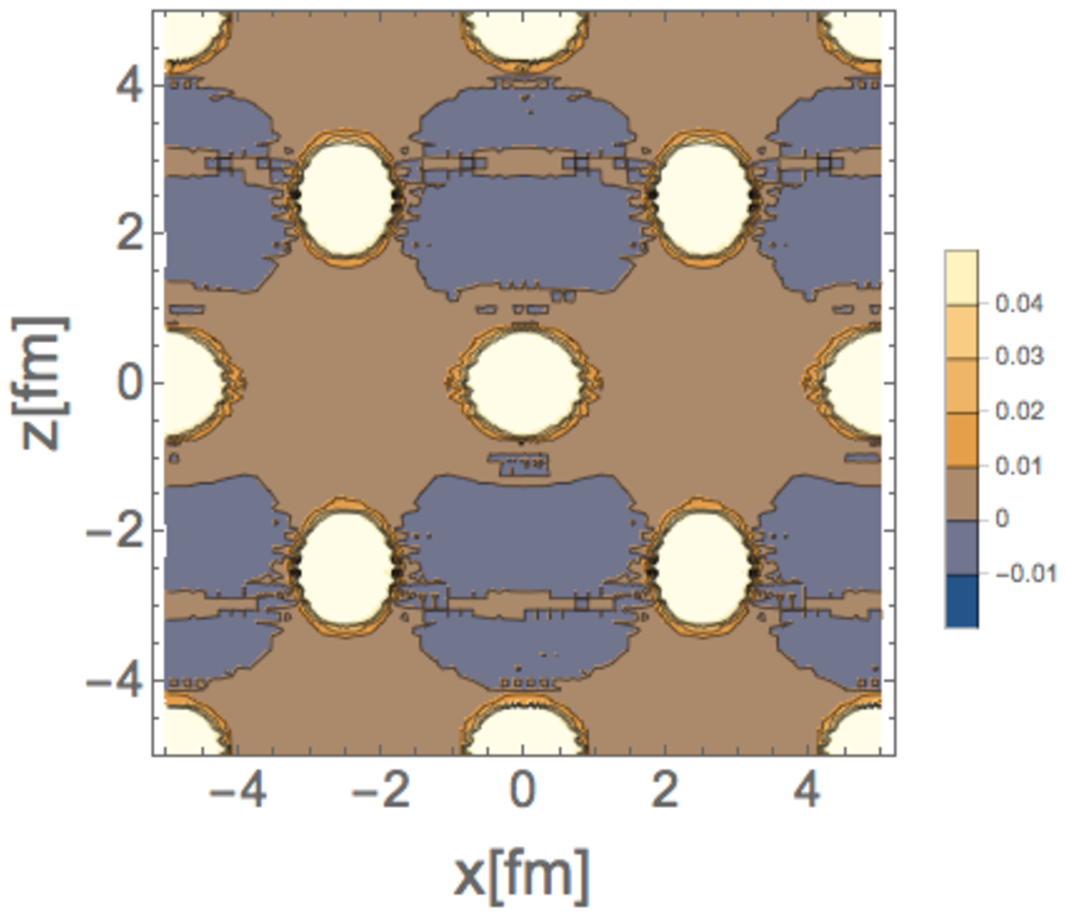}
   \includegraphics[width=5.5cm]{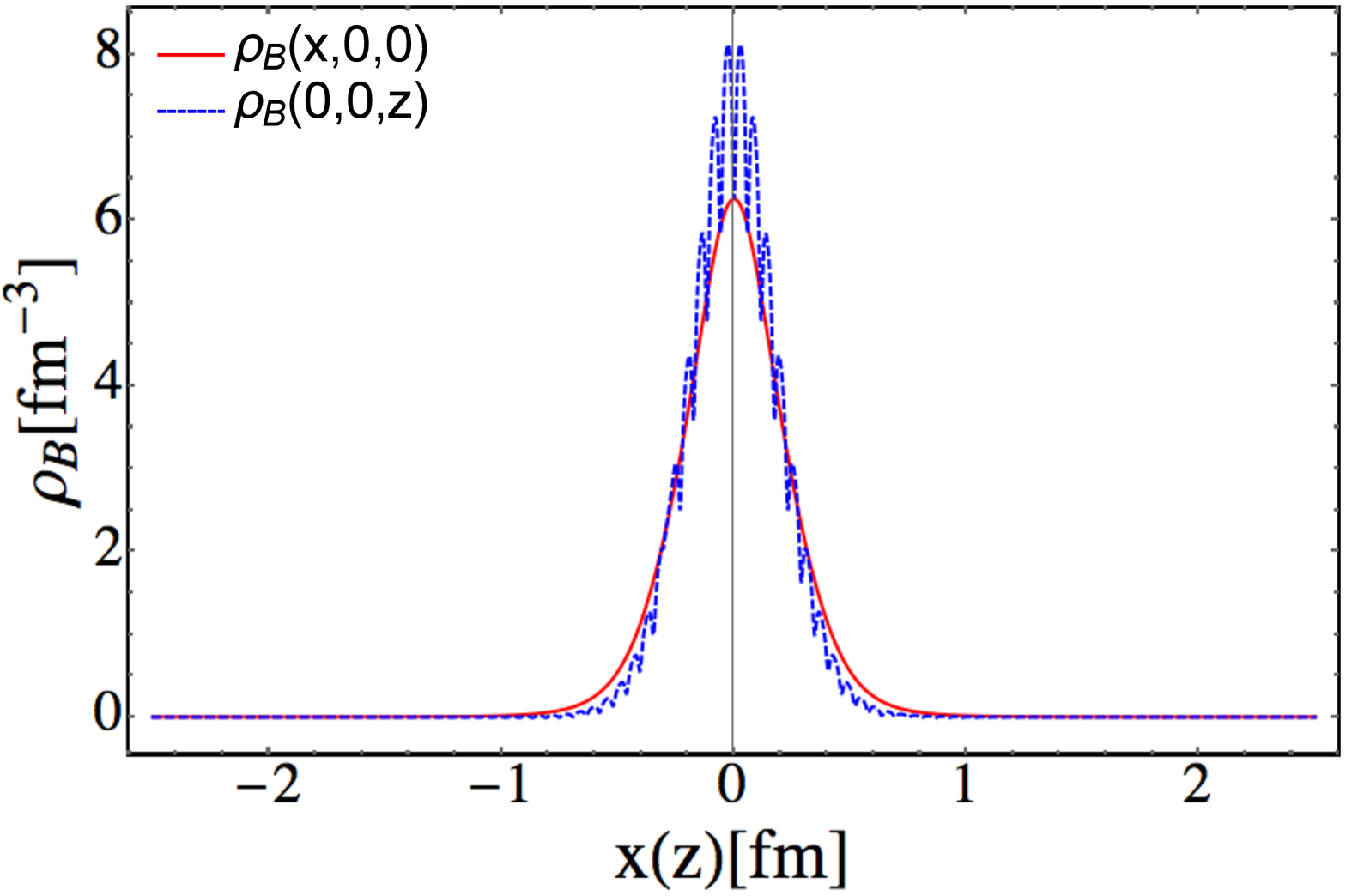}
  \end{center}
 \caption{
  The same as Fig.\ref{bdep25-1} but with $a=0$.
 }
 \label{bdep250}
\end{figure}

\begin{figure}[!htpb]
  \begin{center}
   \includegraphics[width=5.5cm]{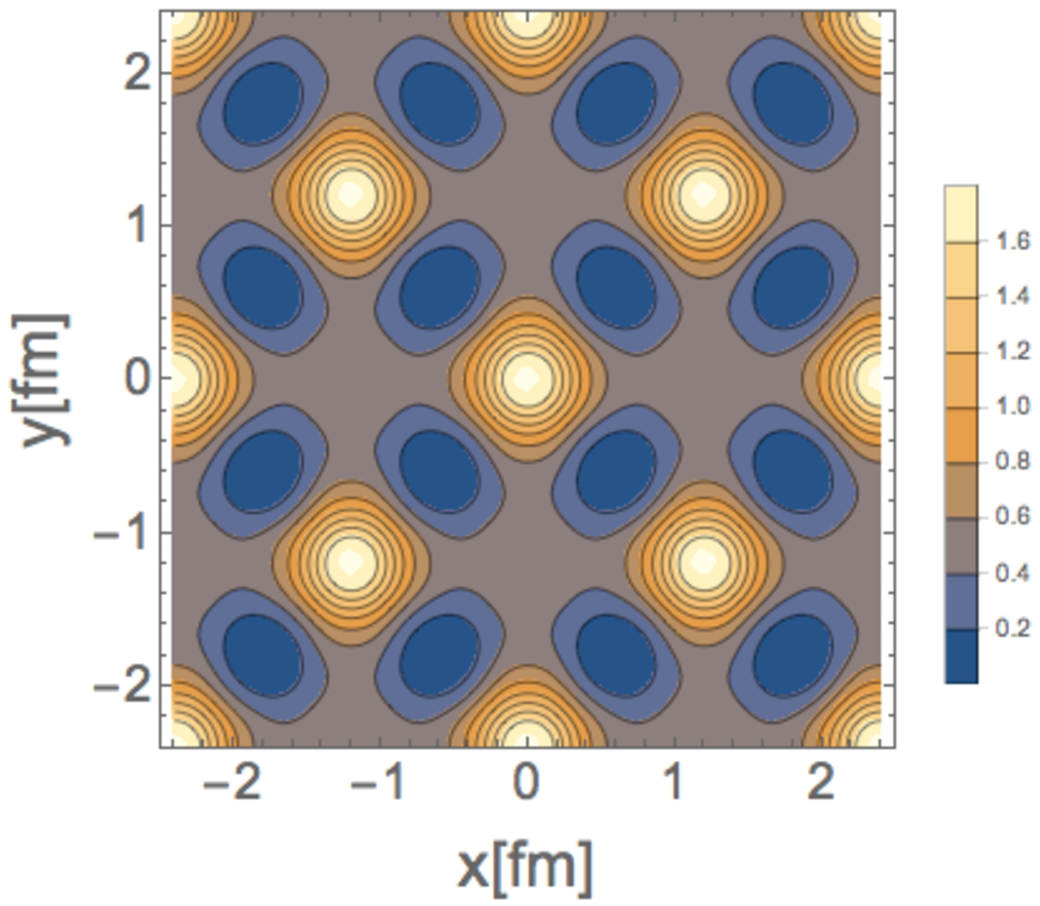}
   \includegraphics[width=5.5cm]{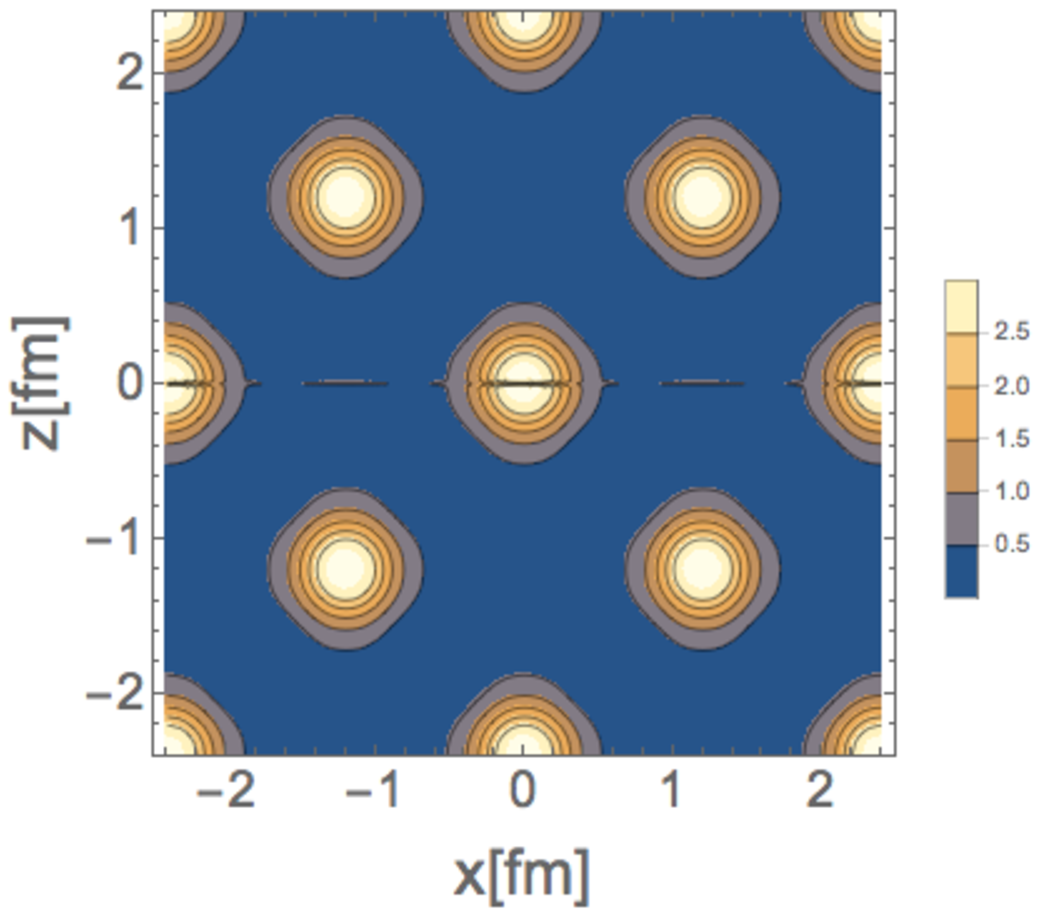}
   \includegraphics[width=5.5cm]{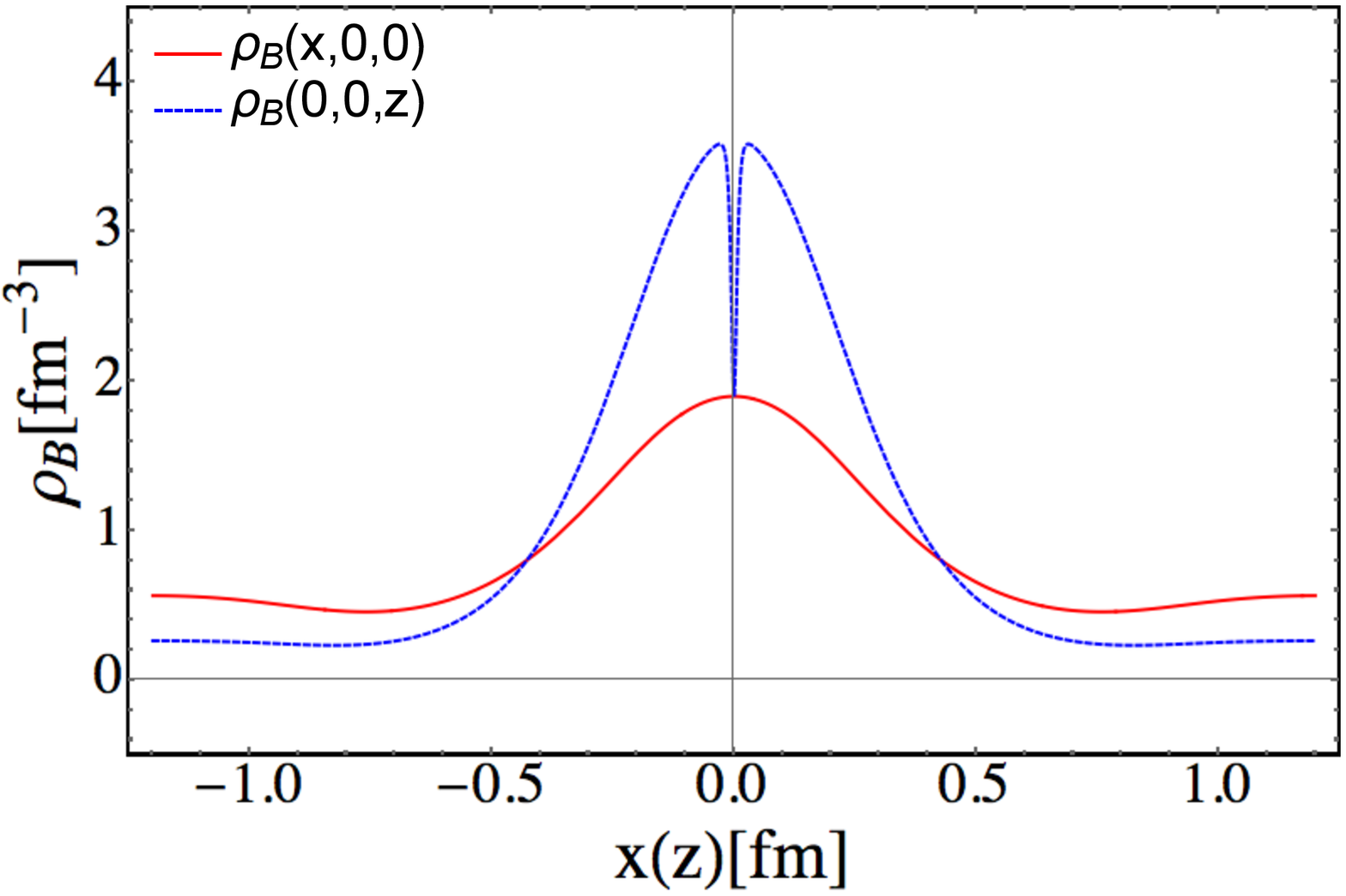}
  \end{center}
 \caption{
  The skyrmion configuration at $L=1.2~{\rm fm}$ and $a=-1$.
 The left(middle) panel displays the density contour plot on x-y plane specified at $z=0$
(x-z plane specified at $y=0$),
 and the lower panel corresponds to the distribution along x-axis or z-axis.
 }
 \label{bdep12-1}
\end{figure}

\begin{figure}[!htpb]
  \begin{center}
   \includegraphics[width=5.5cm]{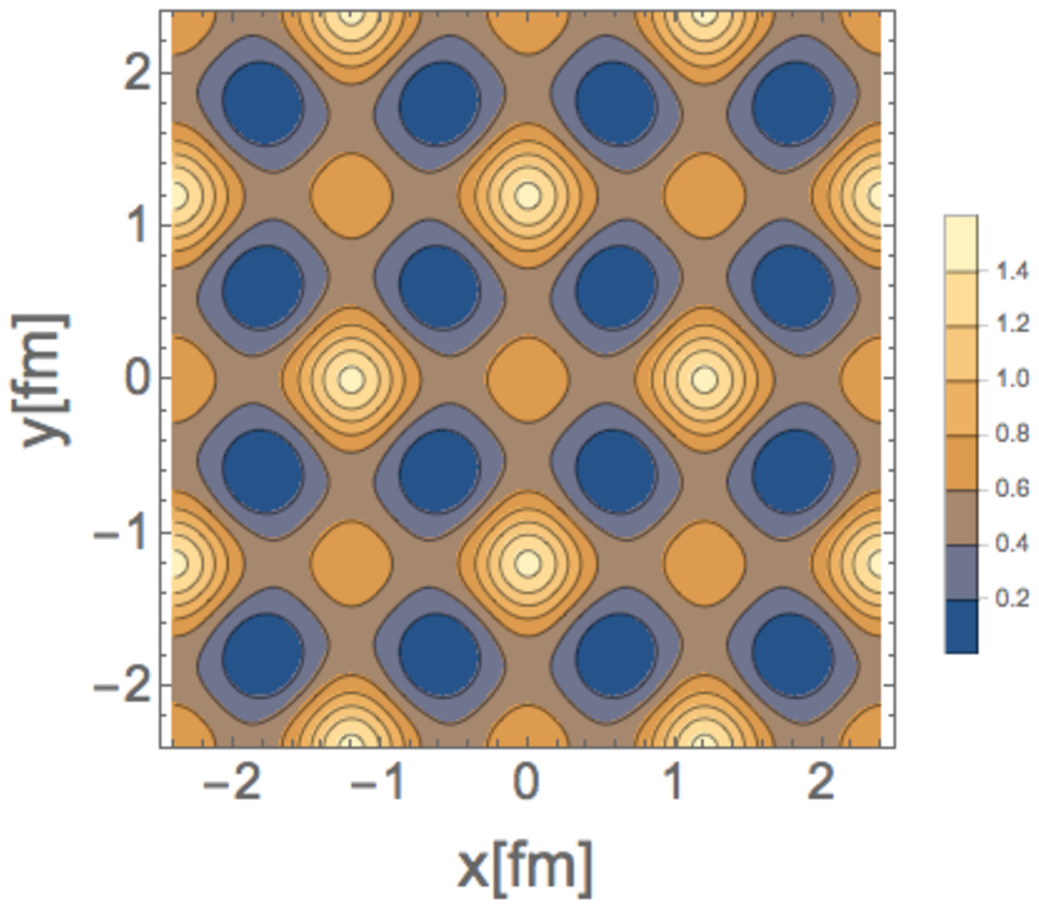}
   \includegraphics[width=5.5cm]{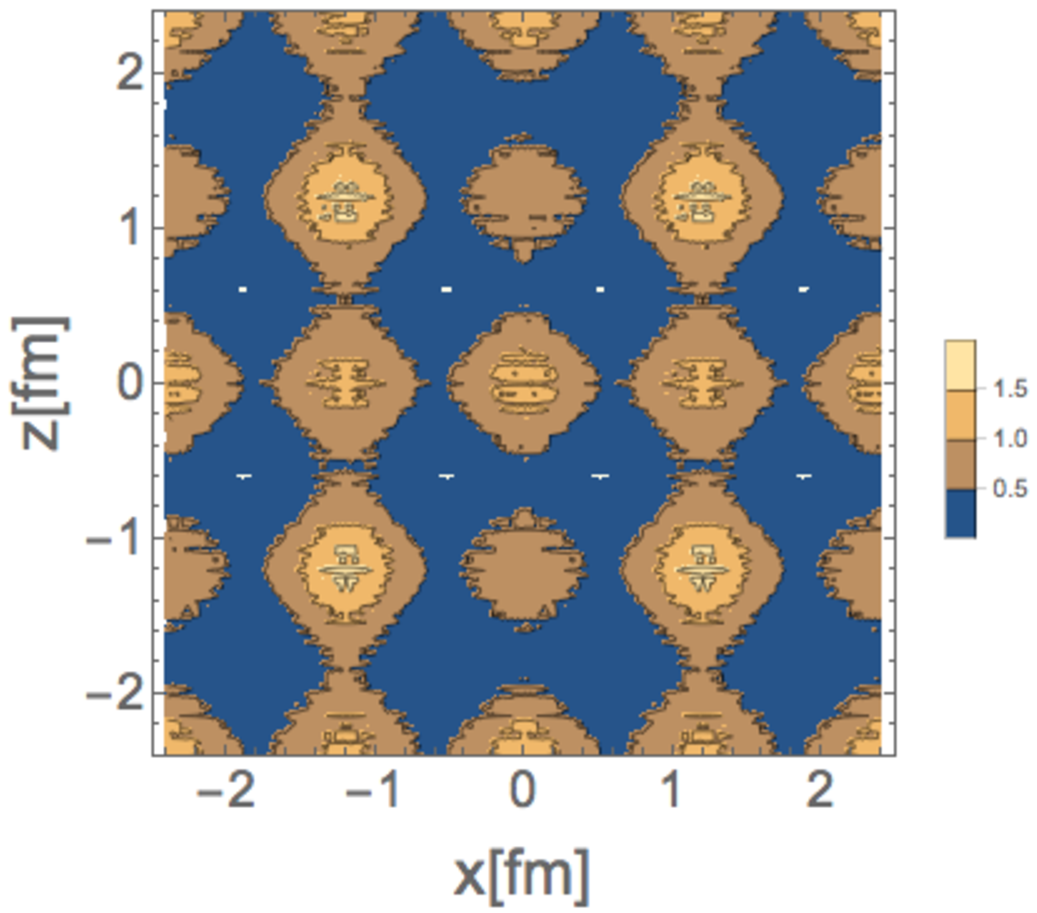}
   \includegraphics[width=5.5cm]{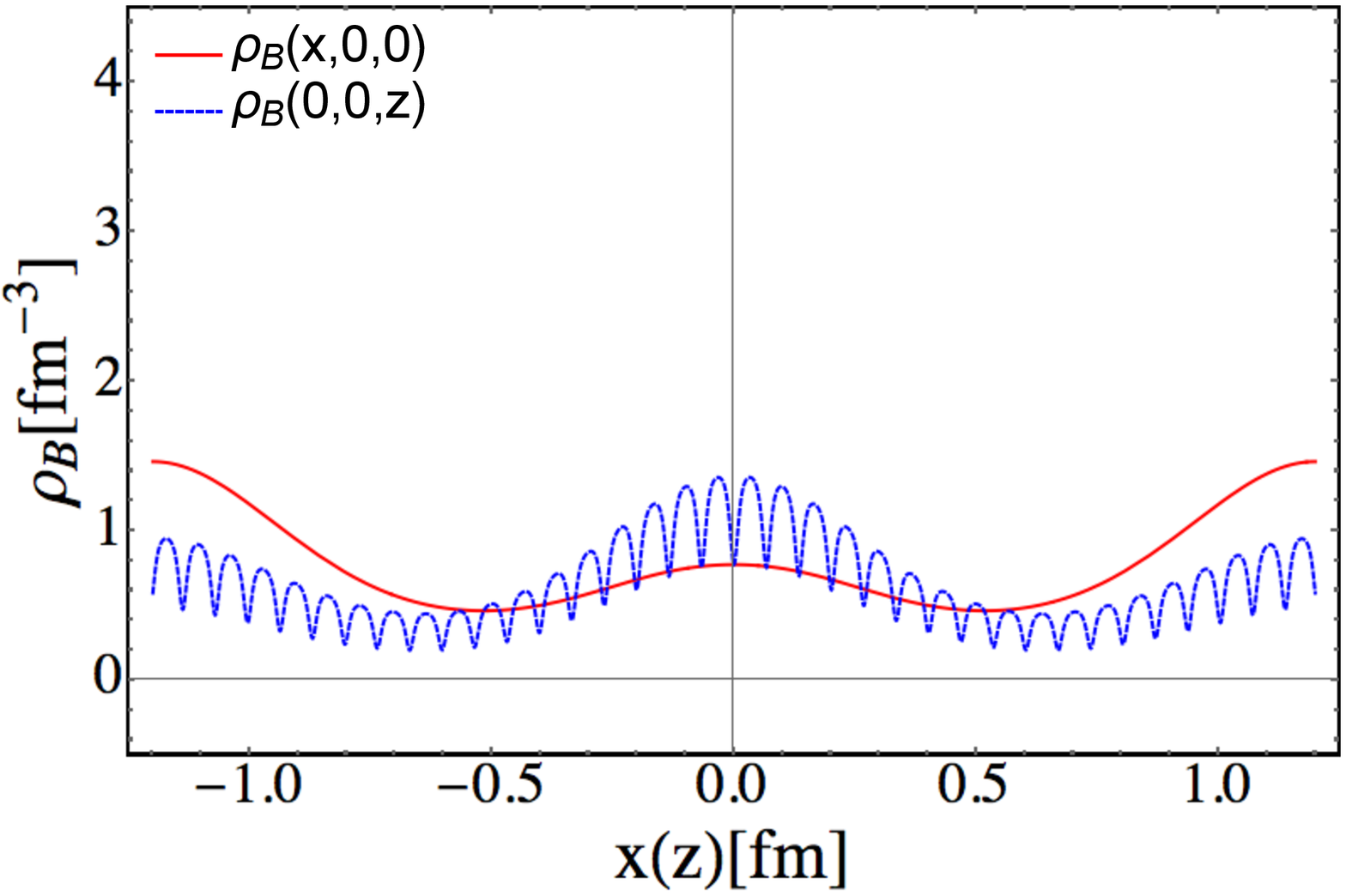}
  \end{center}
 \caption{
 The same as Fig.\ref{bdep12-1}
 but with $a={}-0.1$.
  }
 \label{bdep12-01}
\end{figure}

In Figs.~\ref{bdep25-1}, \ref{bdep25-01} and \ref{bdep250},
we plot the skyrmion configurations in the FCC crystal with $L=2.5~{\rm fm}$ and
varied ($a=-1,-0.1\;{\rm and} \;0$). From these figures, we see that the single baryon shape in the presence of the CSL
is deformed to be higher-intense objects with high frequency.
Particularly, it is interesting to note that as the parameter $a$ increases, the frequency of
the single baryon gets larger by involving the CSL with high frequency
 (equivalently the period gets shorter), as  was seen in the upper panels of  Figs.~\ref{prof_pi1} and \ref{prof_pi3}.
This could trigger the enhancement  of the per-baryon energy with increase of $a$,
as was observed in Fig.~\ref{energy_tot}. Regarding the deformation of skyrmion configuration, the essentially same phenomena  takes place even in the high dense region where $L=1.2~{\rm fm}$ (see  Figs.~\ref{bdep12-1} and~\ref{bdep12-01}) .

\clearpage
\section{Summary}

\label{sum}

In this paper,
we made the first attempt to study the CSL effects on the baryonic matter based on the skrymion crystal approach.
It was found that the CSL plays a role of an inverse catalyzer for the topology change on the baryonic matter modeled  as skyrmoin crystal. We also found
the amplitude of the CSL  becomes small,
in correlation with the pion decay constant in medium,
when the baryon density gets large enough to reach the topology change point.
This could be related to a signal of chiral restoration.
Furthermore, we observed that
the CSL makes the single-baryon shape deformed to be highly oscillating
as the frequency of the CSL gets larger, which leads to the enhancement of the per-baryon energy.
The things we have found in the present paper might be relevant to deeper understanding in condensed-matter systems as well as in compact stars.

%%%%%%%%%%%%%%%%%%%%%%%%%%%%%%%%%%%%%%%%%%%%%%%%%%%%%%%%%%%%%%%%%%%%%%%%%%%

\acknowledgments

The work of M.~K. is supported in part by JSPS Grant- in-Aid for JSPS Research Fellow No. 18J15329.
 Y.~L. M. was supported in part by National Science Foundation of China (NSFC) under Grant No. 11475071, 11747308 and the Seeds Funding of Jilin University. The work of S.~M. was supported in part by
the JSPS Grant-in-Aid for Young Scientists (B) No. 15K17645.

\appendix

\begin{widetext}
\section{The derivation of Eq.~(\ref{prescription})}

Here, we give the derivation of Eq.~(\ref{prescription}).
\begin{eqnarray}
E_{\rm tot} & = &{} -\int_{-\infty}^\infty d^3 x {\cal L} \nonumber\\
& = & {} -N^3\int_{-L}^L d^3 x {\cal L}_{\rm mat}
 -n^3\int_{-l/2}^{l/2} d^3 x \bar{\cal L}_{\rm pion}
 +N^3\int_{-L}^L d^3 x\,
 f_\pi^2m_\pi^2
 \nonumber\\
& = &
N^3\left(\Biggl[-\int_{-L}^L d^3 x {\cal L}_{\rm mat}\Biggl]+
\frac{n^3}{N^3}\Biggl[8\frac{f_\pi^{*2}m_\pi^*}{k} (l)^2E(k)
 +m_\pi^{*2}f_\pi^{*2}\left(1-\frac{2}{k^2}\right)(l)^3 \Biggl]
  +\int_{-L}^L d^3 x\,m_\pi^2f_\pi^2
\right)
 \nonumber\\
& = &
N^3\left(\Biggl[-\int_{-L}^L d^3 x {\cal L}_{\rm mat}\Biggl]+
\frac{2L}{l}\Biggl[8\frac{f_\pi^{*2}m_\pi^*}{k} (2L)^2E(k)\Biggl]
 +m_\pi^{*2}f_\pi^{*2}\left(1-\frac{2}{k^2}\right)(2L)^3
  +\int_{-L}^L d^3 x\,f_\pi^2m_\pi^2
\right)\nonumber\\
& = &
N^3\left(\Biggl[-\int_{-L}^L d^3 x {\cal L}_{\rm mat}\Biggl]+
(2L)^3m_\pi^{*2}f_\pi^{*2}\Biggl[4\frac{E(k)}{k^2K(k)}
 +\left(1-\frac{2}{k^2}\right)\Biggl]
  +\int_{-L}^L d^3 x\,f_\pi^2m_\pi^2
\right)\nonumber\\
& = &
N^3\left(\Biggl[-\int_{-L}^L d^3 x {\cal L}_{\rm mat}\Biggl]+
m_\pi^{2}f_\pi^{2}a
 \left[\int_{-L}^Ld^3x\phi_0\right]
  +\int_{-L}^L d^3 x\,f_\pi^2m_\pi^2
\right)
\nonumber\\
& = & N^3\left(-\int_{-L}^L d^3 x
{\cal L}_{\rm mat}^{\rm(CSL) }\right)
.
\end{eqnarray}

\end{widetext}

\end{document}